%% file: paper.tex
\documentclass[prd,tightenlines,floatfix,showpacs,preprintnumbers,nofootinbib,eqsecnum,superscriptaddress,twocolumn]{revtex4-1}

\usepackage{color}      
\usepackage{amsbsy}     
\usepackage{amsfonts}       
\usepackage{amsmath}        
\usepackage{amssymb}        
\usepackage{color}
\usepackage{wasysym}
\usepackage{multirow}
\usepackage{mciteplus}
\usepackage{psfrag}
\usepackage{slashed}
\usepackage{cancel}
\usepackage{graphicx}

\usepackage{hyperref}

\newcommand{\MO}{{\tt micrOMEGAs}}
\newcommand{\DS}{{\tt DarkSUSY}}
\newcommand{\CHep}{{\tt CalcHEP}}

\newcommand{\FeynCalc}{{\tt FeynCalc}}

\newcommand{\beq}{\begin{equation}}
\newcommand{\eeq}{\end{equation}}
\newcommand{\bea}{\begin{eqnarray}}
\newcommand{\eea}{\end{eqnarray}}

\catcode`\@=11
\font\manfnt=manfnt
\def\Watchout{\@ifnextchar [{\W@tchout}{\W@tchout[1]}}
\def\W@tchout[#1]{{\manfnt\@tempcnta#1\relax%
  \@whilenum\@tempcnta>\z@\do{%
    \char"7F\hskip 0.3em\advance\@tempcnta\m@ne}}}
\let\foo\W@tchout
\def\dubious{\@ifnextchar[{\@dubious}{\@dubious[1]}}

\def\@dubious[#1]{%
  \color{red}\setbox\@tempboxa\hbox{\@W@tchout#1}
  \@tempdima\wd\@tempboxa
  \list{}{\leftmargin\@tempdima}\item[\hbox to 0pt{\hss\@W@tchout#1}]}
\def\@W@tchout#1{\W@tchout[#1]}
\catcode`\@=12

\begin{document}
\preprint{LAPTH-015/14, MS-TP-14-03}

\title{One-loop corrections to gaugino (co-)annihilation into quarks in the MSSM}


\author{B.~Herrmann}
 \email{herrmann@lapth.cnrs.fr}
 \affiliation{
	LAPTh, Universit\'e de Savoie, CNRS, 9 Chemin de Bellevue, B.P.~110, F-74941 Annecy-le-Vieux, France
  }

\author{M.~Klasen}
 \email{michael.klasen@uni-muenster.de}
 \affiliation{
	Institut f\"ur Theoretische Physik, Westf\"alische Wilhelms-Universit\"at M\"unster, Wilhelm-Klemm-Stra{\ss}e 9, D-48149 M\"unster, Germany
  }

\author{K.~Kova\v{r}\'ik}
 \email{karol.kovarik@uni-muenster.de}
 \affiliation{
	Institut f\"ur Theoretische Physik, Westf\"alische Wilhelms-Universit\"at M\"unster, Wilhelm-Klemm-Stra{\ss}e 9, D-48149 M\"unster, Germany
  }
 
 \author{M.~Meinecke}
 \email{mmein\_03@uni-muenster.de}
 \affiliation{
	Institut f\"ur Theoretische Physik, Westf\"alische Wilhelms-Universit\"at M\"unster, Wilhelm-Klemm-Stra{\ss}e 9, D-48149 M\"unster, Germany
  }

\author{P.~Steppeler}
 \email{p\_step04@uni-muenster.de}
 \affiliation{
	Institut f\"ur Theoretische Physik, Westf\"alische Wilhelms-Universit\"at M\"unster, Wilhelm-Klemm-Stra{\ss}e 9, D-48149 M\"unster, Germany
  }

\date{\today}

\begin{abstract}
We present the full $\mathcal{O}(\alpha_s)$ supersymmetric QCD corrections for gaugino annihilation and co-annihilation into light and heavy quarks in the Minimal Supersymmetric Standard Model (MSSM). We demonstrate that these channels are phenomenologically relevant within the so-called phenomenological MSSM. We discuss selected technical details such as the dipole subtraction method in the case of light quarks and the treatment of the bottom quark mass and Yukawa coupling. Numerical results for the (co-)annihilation cross sections and the predicted neutralino relic density are presented. We show that the impact of including the radiative corrections on the cosmologically preferred region of the parameter space is larger than the current experimental uncertainty from Planck data.
\end{abstract}

\pacs{12.38.Bx,12.60.Jv,95.30.Cq,95.35.+d}

\maketitle


\input{intro.tex}

\input{pheno.tex}

\input{analytical.tex}

\input{results.tex}

\input{conclusion.tex}

\acknowledgments

The authors would like to thank J.~Harz and Q.~Le~Boulc'h for fruitful discussions and A.~Pukhov for providing us with the necessary functions to implement our results into the {\MO} code. This work is supported by DAAD/EGIDE, Project No.\ PROCOPE 54366394. The work of M.K. and M.M.\ is supported by the Helmholtz Alliance for Astroparticle Physics. B.H.\ acknowledges support from Campus France, PHC PROCOPE, Project No.\ 26794YC. The work of P.S.\ is supported by the Deutsche Forschungsgemeinschaft, Project No.\ KL1266/5-1.

\appendix
\input{appendix.tex}

\input{bib.tex}

\end{document}

%% file: intro.tex
\section{Introduction}
\label{Intro}

Today there is striking evidence for the existence of a Cold Dark Matter (CDM) component in the universe, coming from a large variety of astronomical observations such as the rotation curves of galaxies, the inner motion of galaxy clusters, and the Cosmic Microwave Background (CMB), to name just a few. The Planck mission \cite{Planck} has measured the CMB with previously unparalleled precision. These measurements, combined with the information from WMAP polarization data at low multipoles \cite{WMAP9}, allow to determine the dark matter relic density of the universe to
\beq
	\label{Planck}
	\Omega_{\mathrm{CDM}}h^2 = 0.1199 \pm 0.0027,
\eeq
where $h$ denotes the present Hubble expansion rate in units of 100 km s$^{-1}$ Mpc$^{-1}$.

The identification of the nature of CDM represents one of the biggest challenges for modern physics. One popular hypothesis is the existence of a new weakly interacting and massive particle (WIMP), which constitutes (at least a part of) the CDM. Besides the lack of direct experimental evidence, the biggest problem of this hypothesis is the fact that the Standard Model of particle physics (SM) does not contain a WIMP, since neutrinos are too light and can only form hot dark matter. This is a strong hint for physics beyond the Standard Model.

A well motivated example for an extension of the SM is the Minimal Supersymmetric Standard Model (MSSM). Under the assumption that a new quantum number, the so-called $R$-parity, is conserved, the lightest supersymmetric particle (LSP) is stable. In many cases the LSP is the lightest of the four neutralinos $\tilde{\chi}_1^0$, which is a mixture of the bino, wino, and two higgsinos, according to 
\begin{equation}
	\tilde{\chi}_1^0 ~=~ Z_{1\tilde{B}} \tilde{B} + Z_{1\tilde{W}} \tilde{W} + Z_{1\tilde{H}_1} \tilde{H}_1 + Z_{1\tilde{H}_2} \tilde{H}_2 \, ,
\end{equation}
and is probably the most studied dark matter candidate. 

The time evolution of the neutralino number density $n_\chi$ is governed by a nonlinear differential equation, the Boltzmann equation \cite{GondoloGelmini}
\beq
	\label{Boltzmann1}
	\frac{\mathrm{d}n_\chi}{\mathrm{d}t} = -3 H n_\chi 
		- \left\langle\sigma_{\mathrm{ann}}v\right\rangle \Big[ n_\chi^2 
		- \left( n_\chi^{\mathrm{eq}} \right)^2 \Big],
\eeq
where the first term on the right-hand side containing the Hubble parameter $H$ stands for the dilution of dark matter due to the expansion of the universe. The second and third term describe the creation and annihilation of neutralinos. Both of these terms are proportional to the thermally averaged annihilation cross section $\left\langle \sigma_{\mathrm{ann}}v \right\rangle$. The creation is also proportional to the number density in thermal equilibrium $n_\chi^{\mathrm{eq}}$, which for temperatures $T \ll m_{\chi}$, $m_{\chi}$ being the lightest neutralino mass, is exponentially suppressed via
\beq
	\label{neq}
	n_\chi^{\mathrm{eq}} \sim \exp \left\{-\frac{m_\chi}{T}\right\}.
\eeq
Therefore the creation rate drops to zero when the universe cools down. At some later point, the expansion of the universe will finally dominate over the annihilation, and the neutralino freezes out asymptotically.

Taking into account the possibility of co-annihilations between the neutralino and the other MSSM particles, the thermally averaged annihilation cross section can be written as \cite{GriestSeckel, EdsjoGondolo}
\beq
	\left\langle \sigma_{\mathrm{ann}}v\right \rangle = 
		\sum_{i,j} \sigma_{ij}v_{ij}\frac{n_i^{\mathrm{eq}}}{n_\chi^{\mathrm{eq}}}\frac{n_j^{\mathrm{eq}}}{n_\chi^{\mathrm{eq}}},
	\label{sigann}
\eeq
where the sum runs over all MSSM particles $i$ and $j$, ordered according to $m_0 = m_{\chi} < m_1 < m_2 < m_3$ etc. 

As can be seen from Eq.\ (\ref{sigann}), co-annihilations can occur not only  if the LSP is involved, but also among several of its possible co-annihilation partners. However, depending on the exact MSSM scenario under consideration, not all of these contributions are numerically relevant. Indeed, by generalizing Eq.\ (\ref{neq}), the ratios of the occurring equilibrium densities are Boltzmann suppressed according to
\beq
	\frac{n_i^{\mathrm{eq}}}{n_\chi^{\mathrm{eq}}} ~\sim~ \exp\left\{ -\frac{m_i-m_\chi}{T} \right\}.
	\label{Suppression}
\eeq
Consequently, only particles whose masses are close to $m_\chi$ can give sizeable contributions. In the MSSM, relevant particles can be light sfermions, in particular staus or stops, or other gauginos. 

Once the Boltzmann equation for the total number density is solved numerically, the relic density is obtained via
\beq
	\Omega_\chi h^2 ~=~ \frac{m_\chi n_\chi}{\rho_{\mathrm{crit}}}.
\eeq
Here, $n_\chi$ is the current neutralino number density after the freeze-out, obtained by solving the Boltzmann equation, and $\rho_{\mathrm{crit}}$ is the critical density of the universe. The theoretical prediction calculated in this way can be compared with the experimental data, i.e.\ the limits given in Eq.\ (\ref{Planck}). This allows to identify the cosmologically preferred regions of the MSSM parameter space. The obtained constraint is complementary to information from collider searches, precision measurements, direct and indirect searches for CDM.

The standard calculation of the relic density is often carried out by a public dark matter code, such as \MO\ \cite{micrOMEGAs} or \DS\ \cite{DarkSusy}. Both of these codes evaluate the (co-)annihilation cross section at an effective tree level, including in particular running coupling constants and quark masses, but no loop diagrams. However, it is well known that higher-order loop corrections may affect the cross section in a sizeable way.

In order to ensure an adequate comparison with the very precise cosmological data, the uncertainties in the theoretical predictions have to be minimized. For a given supersymmetric mass spectrum, the main uncertainty on the particle physics side resides in the calculation of the annihilation cross sections $\sigma_{ij}$, defined in Eq.\ (\ref{sigann}), which govern the annihilation cross section $\sigma_{\mathrm{ann}}$ and thus the relic density $\Omega_{\chi}h^2$. It is the aim of the present work to improve on this point in the context of gaugino\footnote{For clarification we stress that by gaugino we denote all neutralinos and charginos.} (co-)\ annihilation in the MSSM.

The impact of loop corrections on the annihilation cross section and the resulting neutralino relic density has been discussed in several previous analyses. The supersymmetric QCD (SUSY-QCD) corrections to the annihilation of two neutralinos $\tilde{\chi}_1^0$ into third-generation quark-antiquark pairs have been studied in Refs.\ \cite{DMNLO_AFunnel, DMNLO_mSUGRA, DMNLO_NUHM}. The corresponding electroweak corrections have been investigated in Refs.\ \cite{Sloops2007, Sloops2009, Sloops2010}. Further studies are based on effective coupling approaches \cite{Sloops2011, EffCouplings}, including the co-annihilation of a neutralino with a stau. SUSY-QCD corrections to neutralino-stop co-annihilation can be found in Refs.\ \cite{Freitas2007, DMNLO_Stop1, DMNLO_Stop2}. 

These analyses led to the common conclusion that radiative corrections are non-negligible in the context of relic density calculations, as they may influence the resulting theoretical prediction in a sizeable way. In particular, the impact of the corrections is in general larger than the experimental uncertainty of the WMAP or Planck data.

\begin{figure*}
	\includegraphics[scale=0.91]{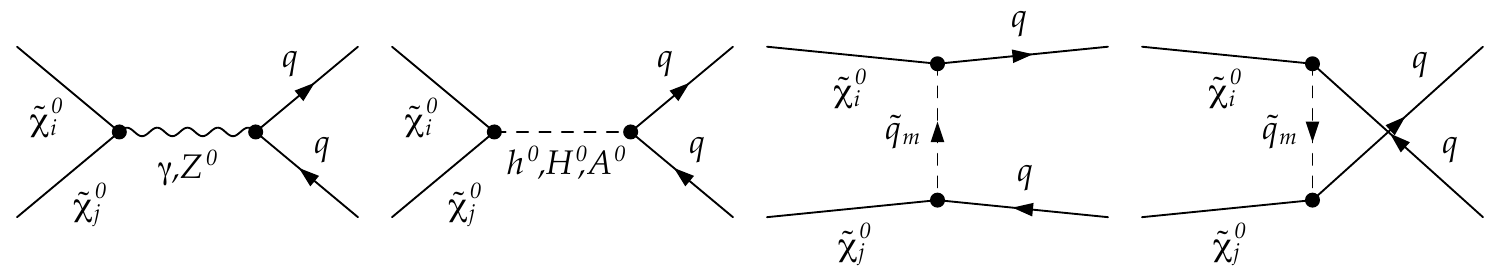}
	\includegraphics[scale=0.91]{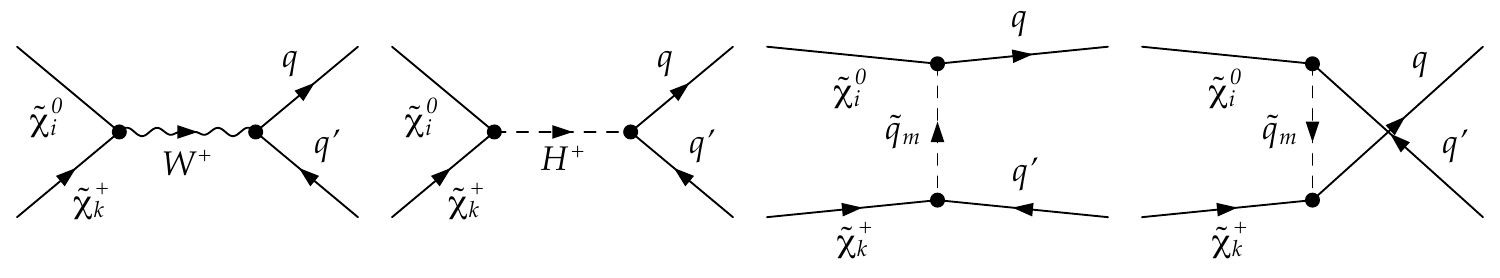}
	\includegraphics[scale=0.91]{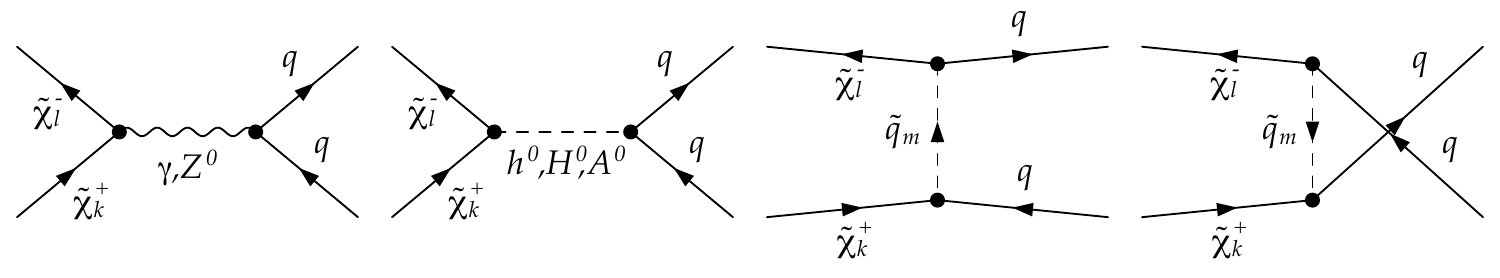}
	\caption{Tree level diagrams of the gaugino (co-)annihilation processes $\tilde{\chi}_i^0 \tilde{\chi}_j^0 \rightarrow q\bar{q}$ (top), $\tilde{\chi}_i^0 \tilde{\chi}_k^\pm \rightarrow q\bar{q}'$ (middle), and $\tilde{\chi}_k^\pm \tilde{\chi}_l^\pm \rightarrow q\bar{q}$ (bottom).}
	\label{TreeLevel}
\end{figure*}

The aim of the present Paper is to extend the calculation of Refs.\ \cite{DMNLO_AFunnel, DMNLO_mSUGRA, DMNLO_NUHM} to all gauginos in the initial and all quarks in the final state. We present the full $\mathcal{O}(\alpha_s)$ corrections in supersymmetric QCD to the following annihilation and co-annihilation processes of gauginos into quark-antiquark pairs:
\bea
	\label{NeuNeuAnni} \tilde{\chi}_i^0   \tilde{\chi}_j^0   & \rightarrow & q\bar{q} ,\\
	\label{NeuChaAnni} \tilde{\chi}_i^0   \tilde{\chi}_k^\pm & \rightarrow & q\bar{q}',\\
	\label{ChaChaAnni} \tilde{\chi}_k^\pm \tilde{\chi}_l^\pm & \rightarrow & q\bar{q}
\eea
for $\{i,j\} = \{1,2,3,4\}$, $\{k,l\} = \{1,2\}$, and $q = \{u,d,c,s,t,b\}$. The quark $q'$ in Eq.\ (\ref{NeuChaAnni}) is the down/up-type quark of the same generation\footnote{In other words, the CKM-matrix is assumed to be diagonal in this analysis.} as the up/down-type quark $q$. The corresponding Feynman diagrams at tree level are shown in Fig.\ \ref{TreeLevel}.

This Paper is organized as follows: In Sec.\ \ref{Pheno} we specify the model framework, introduce our reference scenarios and discuss the phenomenology of gaugino (co-) annihilation. Sec.\ \ref{Technical} contains technical details about the actual cross section calculation. We will discuss the subtleties of the dipole subtraction method for light quarks and the treatment of the bottom quark mass and Yukawa coupling. Aspects concerning the regularization and renormalization are kept rather short, as they can be found in Ref.\ \cite{DMNLO_Stop1}. In Sec.\ \ref{Numerics} we present our numerical results to illustrate the impact of the one-loop corrections on the cross section and the relic density, respectively. Finally, our conclusions are given in Sec.\ \ref{Conclusion}.

%% file: pheno.tex
\section{Phenomenology of gaugino annihilation and co-annihilation}
\label{Pheno}

Throughout this analysis, we work within the phenomenological MSSM (pMSSM)\footnote{Scenarios with important gaugino co-annihilations can, e.g., also be found in models with anomaly mediation \cite{Harigaya:2014dwa}, which are, however, more constrained than our setup.}, where the soft-breaking parameters are fixed at the input scale $Q=1$ TeV according to the SPA convention \cite{SPA}. We choose to work with eleven free parameters, which are detailed in the following: The Higgs sector is fixed by the pole mass of the pseudoscalar Higgs boson $m_{A}$, the higgsino mass parameter $\mu$, and the ratio of the vacuum expectation values of the two Higgs doublets $\tan\beta$. The first and second generation squarks have a common soft-mass parameter $M_{\tilde{q}_{1,2}}$, while the third generation squarks are governed by $M_{\tilde{q}_{3}}$, the soft-mass parameter for the sbottoms and left-handed stops, and $M_{\tilde{u}_{3}}$ for the right-handed stops. All trilinear couplings are set to zero except for $A_t$ in the stop sector. In contrast to the three independent mass parameters in the squark sector, we have a single parameter $M_{\tilde{\ell}}$ for all sleptons. Finally, the gaugino and gluino sector is defined by the bino mass parameter $M_1$, the wino mass parameter $M_2$, and the gluino mass parameter $M_3$. In the context of our analysis, the most interesting parameters are $M_1$, $M_2$, and $\mu$, since they determine the decomposition of the neutralinos and charginos.

Within this setup, with the help of a scan over the parameter space, we have chosen three reference scenarios, which will be used to illustrate the numerical impact of the presented corrections. The corresponding input parameters as discussed above are listed in Tab.\ \ref{ScenarioList}, while Tab.\ \ref{ScenarioProps} summarizes the most important particle masses, mixings, and related observables.

We have used {\tt SPheno~3.2.3} \cite{SPheno} to obtain the physical mass spectrum from the given input parameters. The neutralino relic density and the numerical value of the branching fraction $b \to s \gamma$ have been obtained using {\tt \MO~2.4.1} \cite{micrOMEGAs} with the standard {\tt CalcHEP~2.4.4} \cite{CalcHEP} implementation of the MSSM. The only changes we introduced are that we have set $m_u = m_d = m_s = 0$ as well as included a lower limit on the squark-width, which both do not influence the results concerning dark matter presented here, but will be relevant later in the discussion of the dipole subtraction method in Sec.\ \ref{DipoleSubtraction}. 

Our scenarios have been selected such that they fulfill the following constraints: In order to work with scenarios which are realistic with respect to the recent Planck measurements, we require the neutralino relic density to be in the vicinity of the limits given in Eq.\ (\ref{Planck}). Let us note that we assume that the neutralino accounts for the whole amount of dark matter that is present in our universe. Moreover, we expect the relic density to be modified by our corrections to the (co-)annihilation cross section of the neutralino, so that we apply rather loose bounds at this stage. 

Second, we require the mass of the lightest (``SM-like'') $CP$-even Higgs boson to agree with the observation at LHC,
\begin{equation}
	122~{\rm GeV} \leq m_{h^0} \leq 128~{\rm GeV},
\end{equation}
where we allow for a theoretical uncertainty of about 3 GeV on the value computed by {\tt SPheno}. This uncertainty is motivated by higher-order corrections, which are at present not included in {\tt SPheno}, see, e.g., Ref.\ \cite{Buchmueller:2013psa}. Finally, we impose the interval
\begin{equation}
	2.77\cdot 10^{-4} \leq \mathrm{BR}(b\rightarrow s\gamma) \leq 4.07\cdot 10^{-4}
\end{equation}
on the inclusive branching ratio of the decay $b \to s\gamma$. This corresponds to the latest HFAG value \cite{HFAG} at the $3\sigma$ confidence level. 

\begin{table*}
	\caption{pMSSM input parameters for three selected reference scenarios. All parameters except $\tan\beta$ are given in GeV. }
	\begin{tabular}{|c|ccccccccccc|}
		\hline
			$\quad$ & $\quad\tan\beta\quad$ & $\quad\mu\quad$ & $\quad m_A\quad$ & $\quad M_1\quad$ & $\quad M_2\quad$ & $\quad M_3\quad$ & 		$\quad M_{\tilde{q}_{1,2}}\quad$ & $\quad M_{\tilde{q}_3}\quad$ & $\quad M_{\tilde{u}_3}\quad$ & $\quad M_{\tilde{\ell}}\quad$& $\quad A_t\quad$ \\ 
			\hline 
			I & 13.4 & 1286.3 & 1592.9 & 731.0 & 766.0 & 1906.3 & 3252.6 & 1634.3 & 1054.4 & 3589.6 & -2792.3\\			
			II& 6.6  & 842.3 & 1566.9 & 705.4 & 1928.4 & 1427.0 & 1238.5 & 2352.1 & 774.1 & 2933.2 & -3174.6\\			
			III& 10.0 & 1100.0 & 1951.4 & 1848.0 & 1800.0 & 1102.3 & 3988.5 & 2302.0 & 1636.6 & 1982.1 & -2495.3\\			
			\hline
	\end{tabular}
	\label{ScenarioList}
\end{table*}

\begin{table*}
	\caption{Gaugino masses, the decomposition of the lightest neutralino, and selected observables corresponding to the reference scenarios of Tab.\ \ref{ScenarioList}. All masses are given in GeV.}
	\begin{tabular}{|c|cccc|cc|cccc|ccc|}
		\hline
			$\quad$  & ~~$m_{\tilde{\chi}^0_1}$~~ & ~~$m_{\tilde{\chi}^0_2}$~~ & ~~$m_{\tilde{\chi}^0_3}$~~ & ~~$m_{\tilde{\chi}^0_4}$~~ & ~~$m_{\tilde{\chi}^{\pm}_1}$~~ & ~~$m_{\tilde{\chi}^{\pm}_2}$~~ & ~~$Z_{1\tilde{B}}$~~ &  ~~$Z_{1\tilde{W}}$~~ &  ~~$Z_{1\tilde{H}_1}$~~ &  ~~$Z_{1\tilde{H}_2}$~~ & ~~$m_{h^0}$~~ & ~~$\Omega_{\tilde{\chi}^0_1} h^2$~~ & $\mathrm{BR}(b\rightarrow s\gamma)$ \\
			\hline 
			I  & 738.2 & 802.4 & 1288.4 & 1294.5 & 802.3 & 1295.1 & -0.996 & 0.049 & -0.059 & 0.037 & 126.3 & 0.1243 & $3.0\cdot 10^{-4}$\\			
			II & 698.9 & 850.5 & 854.0 & 1940.2 & 845.6 & 1940.4 & -0.969 & 0.012 & -0.187 & 0.162 &  125.2 & 0.1034 &  $3.2\cdot 10^{-4}$\\			
			III& 1106.7 & 1114.9 & 1855.0 & 1865.6 & 1109.6 & 1856.3 & 0.046 & -0.082 & 0.706 & -0.702 & 126.0  & 0.1190 & $3.2\cdot 10^{-4}$ \\			
			\hline		
	\end{tabular}
	\label{ScenarioProps}
\end{table*}

As can be seen in Tab.\ \ref{ScenarioProps}, the selected scenarios fulfill the mentioned constraints within the required uncertainties. All channels with quark final states contributing to at least 0.1\% of the total annihilation cross section are listed in Tab.\ \ref{ScenarioChannels}, while in Tab.\ \ref{ScenarioSubChannels} we show the contributions of the different sub-channels, i.e.\ the different diagram classes shown in Fig.\ \ref{TreeLevel}. We have grouped the contributions from $s$-channel scalar exchange (contribution denoted $s_S$), the $s$-channel contribution from vector boson exchange ($s_V$), and the squark exchange in the $t$- and $u$-channels ($t/u$). The contributions from the corresponding squared matrix elements are denoted $s_S \times s_S$, $s_V \times s_V$, and $t/u \times t/u$, while the interference terms are denoted by $s_S \times s_V$, $s_S \times t/u$, and $s_V \times t/u$. Note that negative numbers in Tab.\ \ref{ScenarioSubChannels} refer to destructive interferences.

In our scenario I, the dominant contribution to the total annihilation cross section is the co-annihilation between the LSP and the lighter chargino. The second most important channel is the co-annihilation between the two lightest neutralinos, while the pair-annihilation of the LSP is only the third most important channel. This hierarchy is explained as follows: First, as can be seen in Tab.\ \ref{ScenarioSubChannels}, the dominant subchannels for this scenario are the exchange of a scalar in the $s$-channel. More precisely, the value of $\tan\beta = 13.4$ is already large enough to favor bottom quarks in the final states due to the $\tan\beta$-enhanced bottom Yukawa coupling. 

In the case of co-annihilation of the LSP with the second lightest neutralino, this process is mediated by the pseudoscalar Higgs-boson $A^0$, whose mass $m_{A^0} = 1592.9$ GeV is relatively close to the total mass in the initial state, $m_{\tilde{\chi}^0_1} + m_{\tilde{\chi}^0_2} = 1540.6$ GeV. The same argument holds for the co-annihilation with the lighter chargino, which proceeds via the exchange of a charged Higgs boson ($m_{H^{\pm}} = 1595.1$ GeV and $m_{\tilde{\chi}^0_1} + m_{\tilde{\chi}^{\pm}_1} = 1540.5$ GeV). Although these two processes are Boltzmann-suppressed, see Eq.\ (\ref{Suppression}), they are numerically more important than the LSP pair-annihilation, which is kinematically disfavored. Indeed, with $2 m_{\tilde{\chi}^0_1} = 1476.4$ GeV, the configuration is further away from the $A^0$-resonance. 

\begin{table}
\caption{Most relevant gaugino (co-)annihilation channels into quarks in the reference scenarios of Tab.\ \ref{ScenarioList}. Channels which contribute less than 0.1\% to the thermally averaged cross section are not shown.}
	\begin{tabular}{|rl|ccc|}
		\hline
		 & & ~~Scenario I~~ & ~~Scenario II~~ & ~~Scenario III~~\\
		\hline
		$\tilde{\chi}^0_1 \tilde{\chi}^0_1 \to$ & $t\bar{t}$ & 1.4\% & 15.0\% & -- \\
		                                        & $b\bar{b}$ & 9.1\% &  5.9\% & -- \\
		                                        & $c\bar{c}$ & --    &  0.1\% & -- \\
		                                        & $u\bar{u}$ & --    &  0.1\% & -- \\
		\hline
		$\tilde{\chi}^0_1 \tilde{\chi}^0_2 \to$ & $t\bar{t}$ &  2.5\% & 12.0\% & 3.3\% \\
		                                        & $b\bar{b}$ & 23.0\% &  6.9\% & 1.6\% \\
		                                        & $c\bar{c}$ & --     & --     & 1.3\% \\
		                                        & $s\bar{s}$ & --     & --     & 1.7\% \\
		                                        & $u\bar{u}$ & --     & --     & 1.3\% \\
		                                        & $d\bar{d}$ & --     & --     & 1.7\% \\
		\hline
		$\tilde{\chi}^0_1 \tilde{\chi}^0_3 \to$ & $t\bar{t}$ & --  &  9.1\% & -- \\
		                                        & $b\bar{b}$ & --  &  5.3\% & -- \\
		\hline
		$\tilde{\chi}^0_2 \tilde{\chi}^0_2 \to$ & $b\bar{b}$ & 0.2\% & -- & -- \\
		\hline
		$\tilde{\chi}^0_1 \tilde{\chi}^{\pm}_1 \to$ & $t\bar{b}$ & 43.0\% & 40.0\% & 0.8\% \\
													& $c\bar{s}$ & --     & --     & 8.5\% \\
													& $u\bar{d}$ & --     & --     & 8.5\% \\
		\hline
		$\tilde{\chi}^0_2 \tilde{\chi}^{\pm}_1 \to$ & $t\bar{b}$ & 0.4\% & -- & 0.4\% \\
		                                        	& $c\bar{s}$ & 0.9\% & -- & 4.6\% \\
		                                        	& $u\bar{d}$ & 0.9\% & -- & 4.6\% \\
		\hline
		$\tilde{\chi}^{\pm}_1 \tilde{\chi}^{\pm}_1 \to$ & $t\bar{t}$ & 0.2\% & -- & 3.2\% \\
		                                        		& $b\bar{b}$ & 0.6\% & -- & 2.7\% \\
		                                        		& $c\bar{c}$ & 0.2\% & -- & 2.3\% \\
		                                        		& $s\bar{s}$ & 0.2\% & -- & 1.4\% \\
		                                        		& $u\bar{u}$ & 0.2\% & -- & 2.3\% \\
		                                        		& $d\bar{d}$ & 0.2\% & -- & 1.4\% \\
		\hline
		\multicolumn{2}{|c|}{Total} & 83.0\% & 94.4\% & 51.6\% \\
		\hline
	\end{tabular}
	\label{ScenarioChannels}
\end{table}

\begin{table*}
	\caption{Sub-processes for the most important channels of Tab.\ \ref{ScenarioChannels} (more than 2\%) contributing individually at least 0.1\% at $p_{\mathrm{cm}} =100$ GeV.}
	\begin{tabular}{|c|cccccc|}
		\hline
		 & $\quad s_V \times s_V\quad $ &  $\quad s_V \times s_S\quad $ &  $\quad s_S \times s_S\quad $ &  $\quad s_V \times t/u\quad $ &  $\quad s_S \times t/u\quad $ &  $\quad t/u \times t/u\quad $ \\\hline
		 Scenario I &  &  &   &   &  & \\
		$\quad \tilde{\chi}^0_1 \tilde{\chi}^0_1 \to b\bar{b}\quad $      &  --      & --  & 90.5$\%$  &  --       &  9.1$\%$  &  0.4$\%$  \\
                $\quad \tilde{\chi}^0_1 \tilde{\chi}^0_2 \to t\bar{t}\quad $      &  --      & 0.1$\%$  & 27.7$\%$  &  0.1$\%$  &  3.8$\%$  &  33.8$\%$  \\
		$\quad \tilde{\chi}^0_1 \tilde{\chi}^0_2 \to b\bar{b}\quad $      & --      & -- & 96.1$\%$  &  --        &  3.8$\%$   & 0.1$\%$   \\
		$\quad \tilde{\chi}^0_1 \tilde{\chi}^{+}_1 \to t\bar{b}\quad $  &  2.8$\%$ & -- &  79.1$\%$ &   -4.4$\%$ &  11.4$\%$  & 1.1$\%$    \\
		\hline
		 Scenario II &  &  &   &   &  & \\
		$\quad \tilde{\chi}^0_1 \tilde{\chi}^0_1 \to t\bar{t}\quad $      &  --    & --  & 3.2$\%$  &  0.5$\%$  &  1.3$\%$  &  95.0$\%$ \\
		$\quad \tilde{\chi}^0_1 \tilde{\chi}^0_1 \to b\bar{b}\quad $      &  --    & --  & 93.5$\%$  & --  &  6.4$\%$  &  0.1$\%$ \\
		$\quad \tilde{\chi}^0_1 \tilde{\chi}^0_2 \to t\bar{t}\quad $      &  --    &  --  &  91.5 $\%$ & -0.1$\%$ & 7.9$\%$   & 0.7$\%$   \\
                $\quad \tilde{\chi}^0_1 \tilde{\chi}^0_2 \to b\bar{b}\quad $      &  --    &  --  & 99.8$\%$  & -- &  0.2$\%$  &  --  \\
                $\quad \tilde{\chi}^0_1 \tilde{\chi}^0_3 \to t\bar{t}\quad $      &  --    &  --  & 97.8$\%$  & -- & 2.1$\%$  &  0.1$\%$  \\
                $\quad \tilde{\chi}^0_1 \tilde{\chi}^0_3 \to b\bar{b}\quad $      &  --    &  --  & 100.0$\%$  & -- &  --  &  --  \\
		$\quad \tilde{\chi}^0_1 \tilde{\chi}^{+}_1 \to t\bar{b}\quad $  &  0.1$\%$    &  --  &  84.0$\%$ &  -0.5$\%$   & 14.0$\%$   &  2.4$\%$  \\
		\hline
		 Scenario III & &  & &  & & \\
		 $\quad \tilde{\chi}^0_1 \tilde{\chi}^{+}_1 \to c\bar{s}/u\bar{d}\quad $ & 100.4$\%$  & --  & --  &  -0.4$\%$  &  --  &  -- \\
		 $\quad \tilde{\chi}^0_2 \tilde{\chi}^{+}_1 \to c\bar{s}/u\bar{d}\quad $ &  100.0$\%$    & --  & --  &  --  &  --  &  -- \\
                 $\quad \tilde{\chi}^0_1 \tilde{\chi}^0_2 \to t\bar{t}\quad $  & 16.2$\%$    & --  & 1.0$\%$  &  -111.2$\%$  &  -2.7$\%$  & 196.7$\%$ \\
                 $\quad \tilde{\chi}^{+}_1 \tilde{\chi}^{-}_1 \to t\bar{t}\quad $  & 46.2$\%$    & --  & 3.1$\%$  &  -52.9$\%$  &  -4.4$\%$  &  108.0$\%$ \\
                 $\quad \tilde{\chi}^{+}_1 \tilde{\chi}^{-}_1 \to b\bar{b}\quad $  & 21.6$\%$    & --  & 0.7$\%$  &  -131.4$\%$  &  -0.4$\%$  &  209.5$\%$ \\
		 $\quad \tilde{\chi}^{+}_1 \tilde{\chi}^{-}_1 \to c\bar{c}/u\bar{u}\quad $ &  100.2$\%$    & --  & --  &  -0.2$\%$  &  --  &  -- \\
		\hline
	\end{tabular}
	\label{ScenarioSubChannels}
\end{table*}

Finally, although they are kinematically even closer to the $A^0$-resonance ($2 m_{\tilde{\chi}^0_2} \approx 2 m_{\tilde{\chi}^{\pm}_1} \approx 1600$ GeV), the pair annihilation of the lighter chargino or of the second lightest neutralino are highly suppressed by the Boltzmann factor of Eq.\ (\ref{Suppression}) and therefore numerically not relevant.

The main difference in our scenario II is the different setup in the Higgs sector. More precisely, the lower value of $\mu$ together with the higher wino mass modifies the composition of the lightest neutralino such that it has a larger higgsino component. Moreover, due to the smaller value of $\tan\beta=6.6$, down-type Yukawa couplings are less important such that the contribution of final states with top-quarks is larger as can be seen in Tab.\ \ref{ScenarioChannels}. Co-annihilation of the LSP and the lighter chargino (proceeding through $H^{\pm}$ exchange in the $s$-channel) remains the dominant contribution, while co-annihilation with the second lightest neutralino is less relevant in this scenario. This is explained by the different mass spectrum: The mass gap between $\tilde{\chi}^0_1$ and $\tilde{\chi}^{\pm}_1$ is larger than for scenario I, and the mass difference between $\tilde{\chi}^0_1$ and $\tilde{\chi}^0_2$ is even more important, leading to a stronger Boltzmann suppression.

Another difference with respect to scenario I arises in the LSP pair-annihilation. As can be seen in Tab.\ \ref{ScenarioSubChannels}, the dominant subchannel in scenario II is the $t$-channel exchange of a stop. This is due to the fact that the latter is much lighter than in scenario I ($m_{\tilde{t}_1}=874.8$ GeV against $m_{\tilde{t}_1}=1009.0$ GeV in scenario I). For the $\tilde{\chi}^0_1$--$\tilde{\chi}^0_2$ channel, the stop exchange is less relevant, since the kinematical configuration is very close to the resonance of the heavier $CP$-even Higgs boson, $m_{\tilde{\chi}^0_1}+m_{\tilde{\chi}^0_2}=1549.4$ GeV and $m_{H^0} = 1567.1$ GeV.

Finally, scenario II features a non-negligible contribution from co-annihilation of the LSP with the third neutralino, proceeding mainly through scalar exchange in the $s$-channel. As before, this is due to the kinematical situation close to the $H^0$-resonance ($m_{\tilde{\chi}^0_1}+m_{\tilde{\chi}^0_3}=1552.9$ GeV). In both scenarios I and II, the kinematical configuration is such that resonances with $Z^0$- or $W^{\pm}$-exchange are not relevant.

The phenomenology of scenario III is rather different. Here, the lightest neutralino is mainly higgsino-like and the two lightest neutralinos together with the lighter chargino are almost mass-degenerate. Consequently, LSP pair annihilation is negligible, while co-annihilations with the lighter chargino and the second lightest neutralino as well as chargino pair-annihilation are the dominant processes. Moreover, the configuration at this parameter point is such that final states with first and second generation quarks are dominating (contributions from third-generation quarks only amount to 12\%). 

In contrast to the first two reference points, scenario III is characterized by important contributions from the $t$- and $u$-channel diagrams. This is explained by the fact that, here, the lightest stop is relatively light as compared to the annihilating gauginos ($m_{\tilde{t}_1} = 1664.2$ GeV). Consequently, the squark propagator is numerically less suppressed. This configuration also leads to strong destructive interferences between the squark exchange and the $s$-channel contributions. Finally, in this scenario, Higgs exchanges are negligible, since the kinematical configuration is above the corresponding resonances (e.g., $m_{\tilde{\chi}^0_1} + m_{\tilde{\chi}^{\pm}_1} > m_{H^{\pm}} \sim m_{A^0} = 1951.4$ GeV). However, the exchange of a $Z^0$- or $W^{\pm}$-boson in the $s$-channel gives sizeable contributions due to the higgsino nature of the lightest gauginos.

%% file: analytical.tex
\section{Technical details}
\label{Technical}

\begin{figure*}
	\includegraphics[scale=1.0]{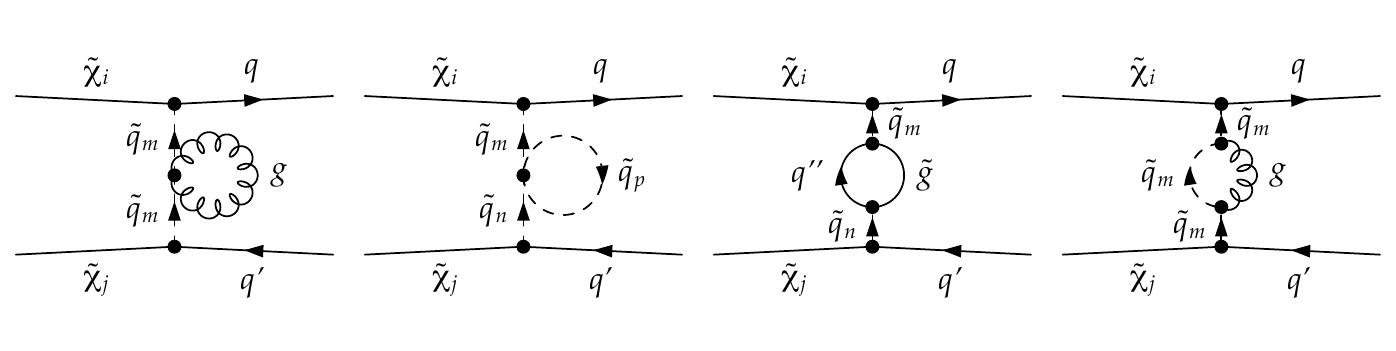}\\[-6mm]
	\includegraphics[scale=1.0]{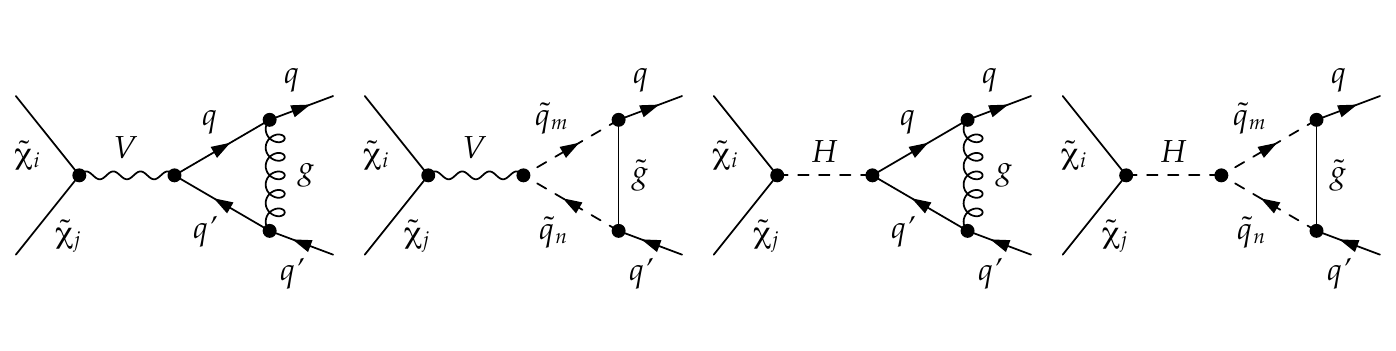}\\[-6mm]
	\includegraphics[scale=1.0]{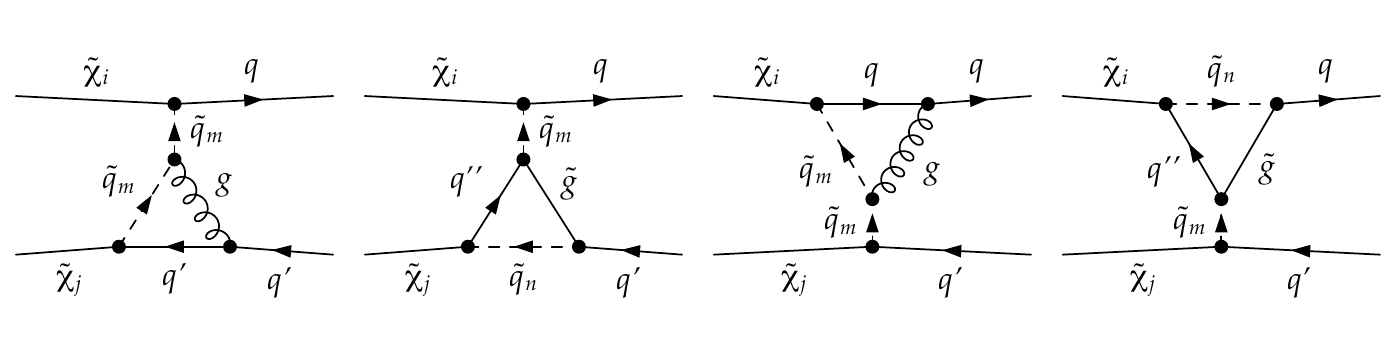}\\[-6mm]
	\includegraphics[scale=1.0]{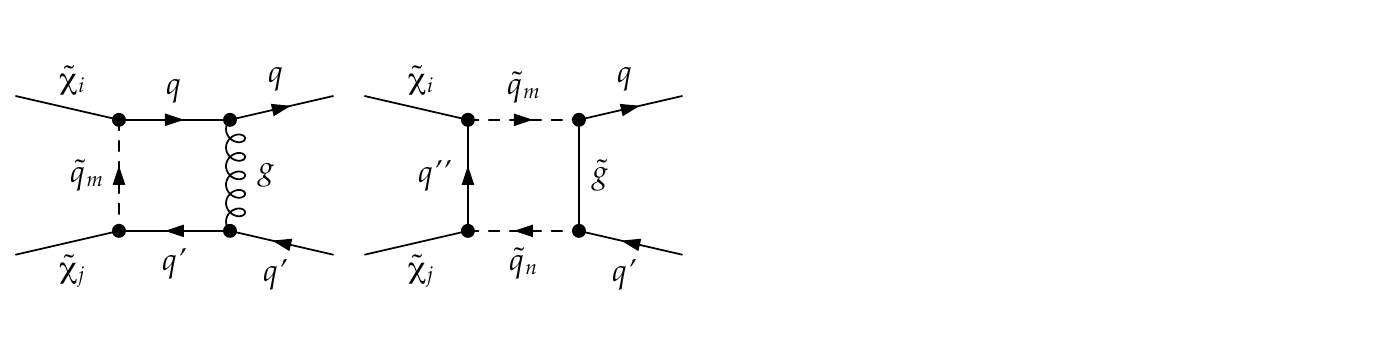}\\[-6mm]
	\caption{Diagrams depicting schematically the one-loop corrections of ${\cal O}(\alpha_s)$ to the gaugino (co-)annihilation processes shown in Fig.\ \ref{TreeLevel}. Here, $V = \gamma, Z^0, W^{\pm}$ and $H = h^0, H^0, A^0, H^{\pm}$. The corrections to the $u$-channel processes are not explicitly shown, as they can be obtained by crossing from the corresponding $t$-channel diagrams.}
	\label{OneLoop}
\end{figure*}

\begin{figure*}
	\includegraphics[scale=1.0]{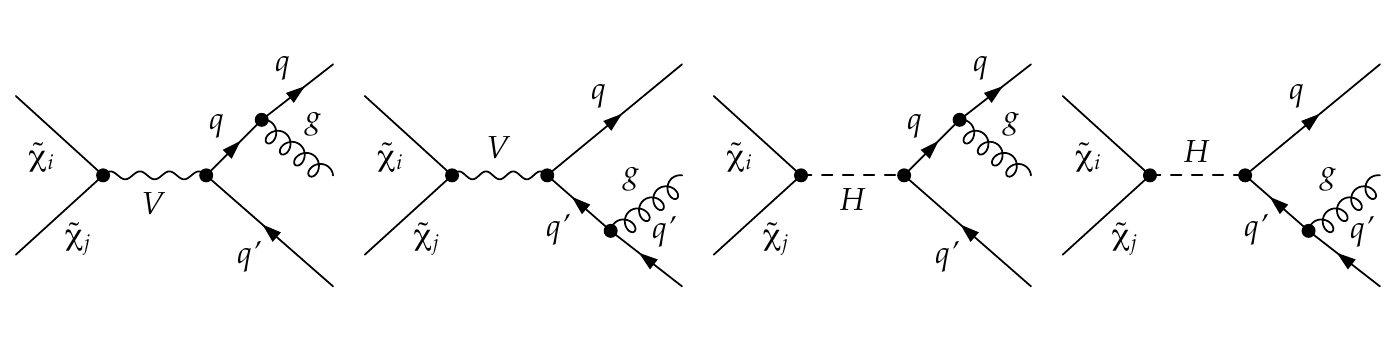}\\[-6mm]
	\includegraphics[scale=1.0]{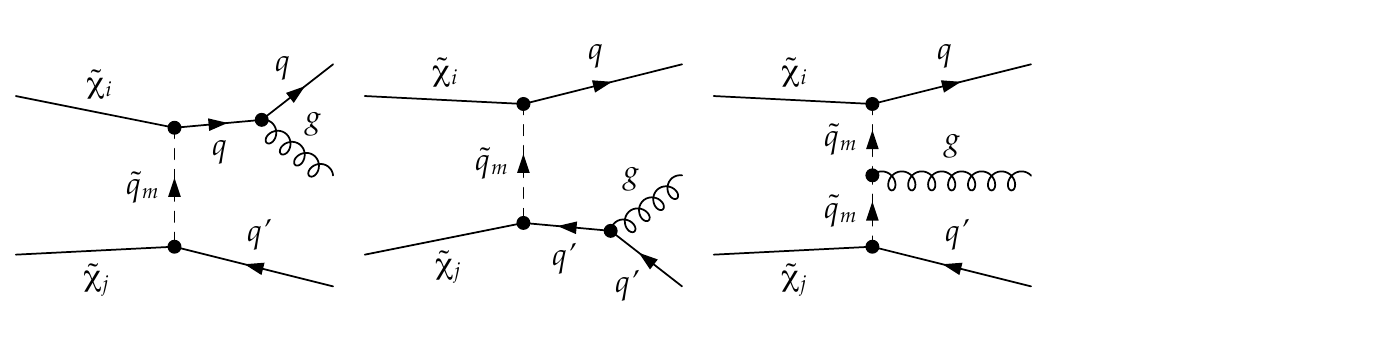}\\[-6mm]
	\caption{Diagrams depicting the real gluon emission corrections of ${\cal O}(\alpha_s)$ to the gaugino (co-)annihilation processes shown in Fig.\ \ref{TreeLevel}. As before, $V = \gamma, Z^0, W^{\pm}$ and $H = h^0, H^0, A^0, H^{\pm}$. The corrections to the $u$-channel processes are not explicitly shown, as they can be obtained by crossing from the corresponding $t$-channel diagrams.}
	\label{RealGluonEmission}
\end{figure*}

The radiative corrections at ${\cal O}(\alpha_s)$ include the one-loop diagrams shown in Fig.\ \ref{OneLoop} as well as the real gluon emission diagrams shown in Fig.\ \ref{RealGluonEmission}. The loop contributions give rise to ultraviolet (UV) and infrared (IR) singularities. While the former are removed via renormalization, i.e.\ the introduction of appropriate counterterms, the latter cancel when including also the real emission of gluons \cite{Dipole:IR-finite}. Altogether, the cross section at next-to-leading order (NLO) in $\alpha_s$ is given by
\begin{equation}
	\label{sigNLO}
	\sigma_{\text{NLO}} ~=~ \int_{2} \mathrm{d}\sigma^{\text{V}} + \int_{3} \mathrm{d}\sigma^{\text{R}},
\end{equation}
where the virtual part ($\mathrm{d}\sigma^{\text{V}}$) and the real emission part ($\mathrm{d}\sigma^{\text{R}}$) are integrated over the two- and three-particle phase space, respectively. We describe the different parts in the following.

\subsection{Calculation of loops}

Here, we focus on the virtual part of the next-to-leading order cross section.
The calculation of the loop diagrams is carried out in the dimensional reduction scheme ($\overline{\rm DR}$), and
therefore all ultraviolet and infrared divergences are regulated dimensionally.
The virtual part of the cross section is rendered UV-finite by redefining the original parameters of the theory.
In our case we introduce counterterms for each of the parameters in the strong sector and specify conditions in order to determine
the counterterms so that all UV divergences vanish. 

In contrast to our previous analyses of neutrino annihilation \cite{DMNLO_mSUGRA, DMNLO_NUHM}, here we use a hybrid on-shell/$\overline{\tt DR}$ renormalization scheme where the parameters $A_{t},A_{b},m^2_{\tilde{t}_1},m^2_{\tilde{b}_1}$,
and $m^2_{\tilde{b}_2}$ are chosen as input parameters along with the heavy quark masses $m_b$ and $m_t$. The trilinear couplings $A_{t},A_{b}$ and the bottom quark mass $m_b$ are defined in the $\overline{\tt DR}$ renormalization scheme, whereas all remaining input masses are defined on-shell. For details of the renormalization scheme see Ref.\ \cite{DMNLO_Stop1}.

In this work we have improved on several aspects of the renormalization scheme. First, we have improved on the determination of the
bottom quark mass in the $\overline{\rm DR}$-scheme. The bottom quark mass in the $\overline{\tt MS}$ renormalization scheme is 
determined from a Standard Model analysis of $\Upsilon$ sum rules \cite{MBmass}. 
The transformation of the $\overline{\tt MS}$ mass $m_b(m_b)$ to the appropriate 
bottom quark mass in the $\overline{\tt DR}$ renormalization scheme within the MSSM requires
several steps as follows:
\begin{widetext}
\begin{equation}
	m_b^{\rm SM,\overline{MS}}(m_b)~\overset{(1)}{\longrightarrow}~ m_b^{\rm SM,
		\overline{MS}}(Q)~\overset{(2)}{\longrightarrow}~ 
	m_b^{\rm SM,\overline{DR}}(Q)~\overset{(3)}{\longrightarrow}~ m_b^{\rm MSSM,\overline{DR}}(Q)\,.
\end{equation}
In the first step we use the three-loop renormalization group evolution to obtain the mass of the bottom quark at a 
scale $Q$ \cite{Chetyrkin:1997dh} \footnote{We use the scale $Q = 1$ TeV for the following numerical analysis.}. Then in the second step at the final scale, while still considering only the Standard Model, we transform the bottom quark mass from the $\overline{\tt MS}$ to the $\overline{\tt DR}$ renormalization scheme using
the two-loop relation \cite{Harlander:2006rj}
\begin{eqnarray}
m_b^{\rm SM,\overline{DR}}(Q)~ &=&~ m_b^{\rm SM,\overline{MS}}(Q) \Bigg[1 -\frac{\alpha_e}{\pi}\frac{1}{4} C_F +
	  \left(\frac{\alpha_s^{\rm \overline{MS}}}{\pi}\right)^{\! 2} \frac{11}{192} C_A C_F - 
	\frac{\alpha_s^{\rm \overline{MS}}}{\pi}\frac{\alpha_e}{\pi}
	  \left(\frac{1}{4} C_F^2 + \frac{3}{32}  C_A C_F \right) 
	  \nonumber\\
	&&\hspace{2cm}+ \left(\frac{\alpha_e}{\pi}\right)^2 \left( \frac{3}{32} C_F^2 
	      +  \frac{1}{32} C_F T n_f\right)   
	      + \ldots
	\Bigg] \,.
\end{eqnarray}
This transformation involves an evanescent coupling $\alpha_e$, which is identical to the strong coupling constant 
$\alpha_s^{\rm \overline{DR}}$ in a supersymmetric theory, whereas in QCD there is a subtle difference. The dots indicate higher order terms, which are not relevant here. The constants $C_A$ and $C_F$ are the usual color factors of QCD, $C_A=3$ and $C_F=4/3$, respectively. The difference between the couplings to one-loop order is \cite{Bauer:2008bj}
\begin{eqnarray}
	\alpha_e &=&
	\alpha_s^{\rm{ \overline{DR}}}\Bigg[ 1+ \frac{\alpha_s^{\rm{ \overline{DR}}}}{ \pi } \Bigg\{-T_F
	 \frac{ L_t}{2}  +\frac{C_A}{4} \Bigg(   
	2+L_{\tilde{g}} + \sum_{i=1,2}\left(L_{\tilde{g}}-L_{\tilde{q}_i}\right)
	\frac{m_{\tilde{q}_i}^2}{m_{\tilde{g}}^2-m_{\tilde{q}_1}^2}\Bigg) \nonumber \\
	&&\hspace{1cm} + \frac{C_F}{4} \Bigg( 
	\sum_{i=1,2} \left(-1-2 L_{\tilde{g}}+2 L_{\tilde{q}_i}
	+\left(-L_{\tilde{g}}+L_{\tilde{q}_i}\right) 
	\frac{m_{\tilde{q}_i}^2}{m_{\tilde{g}}^2 - m_{\tilde{q}_i}^2} \right) \frac{
	  m_{\tilde{q}_i}^2}{ m_{\tilde{g}}^2 -m_{\tilde{q}_i}^2} + \left(-3-2 L_{\tilde{g}}\right) \Bigg) \Bigg\} \Bigg],
\end{eqnarray}
\end{widetext}	
where $\alpha_s^{\rm{ \overline{DR}}}$ stands for the strong coupling constant in the $\overline{\tt DR}$ renormalization scheme in the MSSM and we have used the shorthand notation $L_i = \log(Q^2/m_i^2)$. 

The last step requires adding threshold corrections from supersymmetric particles 
in the loop. Using the results of Ref.\ \cite{Bauer:2008bj}, we can write the final transformation as
\begin{equation}\label{mbsmmssm}
m_b^{\rm SM,\overline{DR}}(Q) ~=~ \zeta_{m_b}\, m_b^{\rm MSSM,\overline{DR}}(Q)\,, 	
\end{equation}
where the coefficient $\zeta_{m_b}$ can be expanded in the strong coupling constant,
\begin{equation}
	\zeta_{m_b} ~=~ 1 + \Biggr( \frac{\alpha_s^{\rm{ \overline{DR}}}}{ \pi }\Biggr) \, \zeta^{(1)}_{b} + 
	\Biggr( \frac{\alpha_s^{\rm{ \overline{DR}}}}{ \pi } \Biggr)^{\! 2} \zeta^{(2)}_{b} + \mathcal{O}(\alpha_s^3) \, .
\end{equation}
Using the results for the coefficients $\zeta^{(1)}_{b}$ and $\zeta^{(2)}_{b}$ given in Ref.\ \cite{Bauer:2008bj} and 
inverting Eq.~(\ref{mbsmmssm}) yields the final bottom quark mass in the $\overline{\tt DR}$ renormalization scheme
in the MSSM which is subsequently used in our analysis.

The bottom quark mass deserves our special attention as the Higgs exchange is the leading contribution to many of the
(co-)annihilation cross sections in this analysis. Therefore the second improvement of this study with respect to our previous 
works involves an improvement to the Yukawa coupling of the bottom quark. Similar to our earlier analyses, we improve our 
full one-loop SUSY-QCD by including higher-order QCD and SUSY-QCD corrections to the Yukawa coupling.

Leading QCD and top-quark induced corrections to the Yukawa coupling of Higgs bosons to bottom quarks were calculated up to 
$\mathcal{O}(\alpha_s^4)$ \cite{QCDhiggs} and can be used to define an effective Yukawa coupling which
includes these corrections as
\begin{equation}
	\Big[ h_b^{\overline{\tt MS},{\tt QCD},\Phi} (Q)\Big]^2 = 
	\Big[ h_b^{\overline{\tt MS},\Phi} (Q)\Big]^2 \Big[1 + \Delta_{\tt QCD} + \Delta_{t}^\Phi\Big]\,,
\end{equation}
for each Higgs boson $\Phi=h^0,H^0,A^0$. The QCD corrections $\Delta_{\tt QCD}$ are explicitly given by
\begin{align}\nonumber
	\Delta_{\tt QCD} &\ \! = \frac{\alpha_s(Q)}{\pi} C_F\frac{17}{4} + \frac{\alpha^2_s(Q)}{\pi^2}\Big[ 35.94 - 1.359\, n_f \Big]\\
	& + \frac{\alpha^3_s(Q)}{\pi^3}\Big[164.14 - 25.77\, n_f + 0.259\,n^2_f\Big]\\ \nonumber & + 
	\frac{\alpha^4_s(Q)}{\pi^4}\Big[39.34 - 220.9\,n_f +9.685\,n^2_f - 0.0205\, n^3_f\Big]\,,
\end{align}
and the top-quark induced corrections $\Delta_{t}^\Phi$ for each Higgs boson $\Phi$ read
\begin{align}
	\Delta_{t}^h = c_h(Q)&\Biggr[ 1.57 - \frac23 \log \frac{Q^2}{m_t^2}+ 
	\frac19 \log^2\frac {m^2_b(Q)}{Q^2} \Biggr],\\
	\Delta_{t}^H = c_H(Q)&\Biggr[ 1.57 - \frac23 \log \frac{Q^2}{m_t^2}+ 
	\frac19 \log^2\frac {m^2_b(Q)}{Q^2} \Biggr], \\
	\Delta_{t}^A = c_A(Q)&\Biggr[ \frac{23}{6} - \log \frac{Q^2}{m_t^2}+ 
	\frac16 \log^2\frac {m^2_b(Q)}{Q^2} \Biggr],
\end{align}
with
\begin{multline}
	\Big\{ c_h(Q),~ c_H(Q),~ c_A(Q) \Big\} = \\ \frac{\alpha_s^2(Q)}{\pi^2}
	\Biggr\{ \frac{1}{\tan\alpha\tan\beta}, ~ 
	      \frac{\tan\alpha}{\tan\beta}, ~ 
	      \frac{1}{\tan^2\beta} \Biggr\}\,.
\end{multline}
We exclude the one-loop part of these corrections as it is 
provided consistently through our own calculation.

In the MSSM, the Yukawa coupling to bottom quarks can be enhanced for 
large $\tan\beta$ or large $A_b$, and then effects even beyond the next-to-leading order should be included. 
Therefore, in addition, we incorporate these corrections that can be resummed 
to all orders in perturbation theory \cite{Carena2000,Spira2003}. Denoting the 
resummable part by $\Delta_b$, we redefine the bottom quark Yukawa couplings as
\begin{eqnarray}
	h_b^{{\tt MSSM},h}(Q) &=& \frac{h_b^{\overline{\tt MS},{\tt QCD},h}(Q)}{1+\Delta_b}
	\Biggr[1-\frac{\Delta_b}{\tan\alpha\tan\beta}\Biggr]\,, \qquad \quad \\
	h_b^{{\tt MSSM},H}(Q) &=& \frac{h_b^{\overline{\tt MS},{\tt QCD},H}(Q)}{1+\Delta_b}
	\Biggr[1+\Delta_b \frac{\tan\alpha}{\tan\beta}\Biggr]\,, \\
	h_b^{{\tt MSSM},A}(Q) &=& \frac{h_b^{\overline{\tt MS},{\tt QCD},A}(Q)}{1+\Delta_b}
	\Biggr[1-\frac{\Delta_b}{\tan^2\beta}\Biggr]\,.
\end{eqnarray}
As a further improvement, we include also the leading NNLO contributions to $\Delta_b$
as given in Ref.\ \cite{Noth:2010jy}. These corrections are now also available for general, i.e.\ non-minimal sources of flavour violation \cite{Crivellin:2012zz}. The electroweak one-loop corrections have also been computed and resummed analytically to all orders \cite{Crivellin:2011jt}. We leave their implementation to future work.

Let us finally briefly comment on possible Sommerfeld enhancement
effects of gaugino \mbox{(co-)}annihilation in the MSSM. As is well known,
potentially large loop corrections arise from long-range interactions
of WIMPs before their annihilation, which are mediated by bosons with
masses $m_\phi$ well below the WIMP mass $m_\chi$. More precisely, these
corrections become relevant when the Bohr radius $1/(\alpha_W m_\chi)$
is smaller than the interaction range $1/m_\phi$ or $\alpha_W m_\chi/m_\phi
\geq 1$. This condition is almost never realized in the MSSM with an
LSP mass of $m_\chi \leq 1$ TeV, $\alpha_W \approx 1/30$ and weak gauge and Higgs
boson masses $m_\phi \sim {\cal O}(100)$ GeV \cite{Drees:2013er}. In
particular, it is never realized in our scenarios I--III. Note, however,
that in scenario III the pair annihilation of charginos, which are
almost mass degenerate with the LSP, contributes about 10\% to the total
cross section, so that this (subleading) channel would indeed be enhanced
through multiple exchanges of massless photons if resummed to all orders
\cite{Beneke:2012tg, Hellmann:2013jxa}. These calculations, which involve
either numerical solutions of coupled Schr\"odinger equations or analytic
calculations in an effective non-relativistic theory, are postponed to
future work.

\subsection{Dipole subtraction method}
\label{DipoleSubtraction}

As mentioned above, the real gluon emission shown in Fig.\ \ref{RealGluonEmission} needs to be included in order to cancel the remaining infrared (IR) singularities in the virtual part of the cross section \cite{Dipole:IR-finite}. However, this is not as straightforward as in the ultraviolet case, since the two contributions reside in the differential cross sections $\mathrm{d}\sigma^{\text{V}}$ and $\mathrm{d}\sigma^{\text{R}}$, which are integrated over different phase spaces. Moreover, working in $D=4-2\epsilon$ dimensions, the soft and collinear divergencies appearing in the virtual contribution can be explicitly isolated and appear as single and double poles, $1/\epsilon$ and $1/\epsilon^2$,
while the divergencies in the real corrections arise from the phase space integration over the gluon phase space. In addition, quasi-collinear divergencies can appear in $\sigma^{\text{R}}$ including large logarithmic corrections of the form $\log(s/m^2)$, which cancel against logarithms of the same form in $\sigma^{\text{V}}$. 

For these reasons, and generally speaking, a separate numerical evaluation of the two phase-space integrations in Eq.\ (\ref{sigNLO}) cannot lead to numerically stable results.
There are two approaches to render both of these terms separately infrared and collinear safe and therefore numerically evaluable: The so-called phase-space slicing method \cite{Dipole:phase-space-slicing} and the dipole subtraction method {\cite{Dipole:Dipole0, Dipole:Dipole1, Dipole:Dipole2}. In the present work, we shall use the latter, which we will describe in the following.

The dipole subtraction method renders the integrands in Eq.\ (\ref{sigNLO}) seperately finite by adding and 
subtracting an auxiliary cross section $\mathrm{d}\sigma^{\text{A}}$. Using dimensional regularization, this is done according to
\begin{equation}
 	\label{Dipole0}
 	\sigma_{\text{NLO}} \!= \!\! \int_{3} \! \biggr[ \mathrm{d}\sigma^{\text{R}} \Big|_{\epsilon=0} - 	
		\mathrm{d}\sigma^{\text{A}} \Big|_{\epsilon=0} \biggr]\! 
		+\!\! \int_2 \! \left[ \mathrm{d}\sigma^{\text{V}} + 
		\!\int_1 \! \mathrm{d}\sigma^{\text{A}} \right]_{\epsilon=0}\!\!,
\end{equation}
where in the last term on the right-hand side the three-particle phase-space integral is factorized into the two-particle phase-space integral of $\sigma^{\text{V}}$ and the integration over the one-particle phase-space of the radiated gluon. 
The auxiliary cross section $\mathrm{d}\sigma^{\text{A}}$, acting as a local counterterm for $\mathrm{d}\sigma^{\text{R}}$, has to possess the same pointwise singular behavior as $\mathrm{d}\sigma^{\text{R}}$ and has to be analytically integrable over the gluon phase space in $D$ dimensions. 
Then, on the one hand, $\mathrm{d}\sigma^{\text{A}}$ reproduces the potentially soft or collinear singular terms in the real corrections, such that one ends up with a convenient form for numerically performing the three-particle phase-space integration in Eq.\ (\ref{Dipole0}). On the other hand, $\int_1 \mathrm{d}\sigma^{\text{A}}$ cancels all single and double poles appearing in $\mathrm{d}\sigma^{\text{V}}$ in a way that the sum $\mathrm{d}\sigma^{\text{V}}+\int_1 \mathrm{d}\sigma^{\text{A}}$ is rendered finite even in the limit $D \rightarrow 4$. In addition, $\mathrm{d}\sigma^{\text{A}}$ can be written in such a way, that it also cancels all quasi-collinear divergencies.

The dipole contributions to the matrix elements $|M^{\text{R}}|^2$ of real corrections in the case of final state radiation can be written in the general form
\begin{equation}
 	\label{Dipole1}
  	\big| M^{\text{R}} \big|^2=\sum_{i,j} \sum_{k\neq i,j} \mathcal{D}_{ijk} + \dots = 
		\mathcal{D}_{gq,\bar{q}} + \mathcal{D}_{g\bar{q},q} + \dots 
		\hspace{3mm} .
\end{equation}
This expression encodes the singular structure of the real radiation matrix element as a summation over so-called emitter-spectator pairs, singled out over the two Born-level external particles in all possible ways, and the dots stand for further infrared and collinear finite terms. 
Here, $i$ and $j$ run over the final state particles connected to the emitter through a splitting process as depicted in Fig.\ \ref{Dipole:Bild2}, and $k$ stands for the spectator particle, which is needed to maintain conservation of gauge-group charges and total momentum.

\begin{figure}
	\includegraphics[scale=0.55]{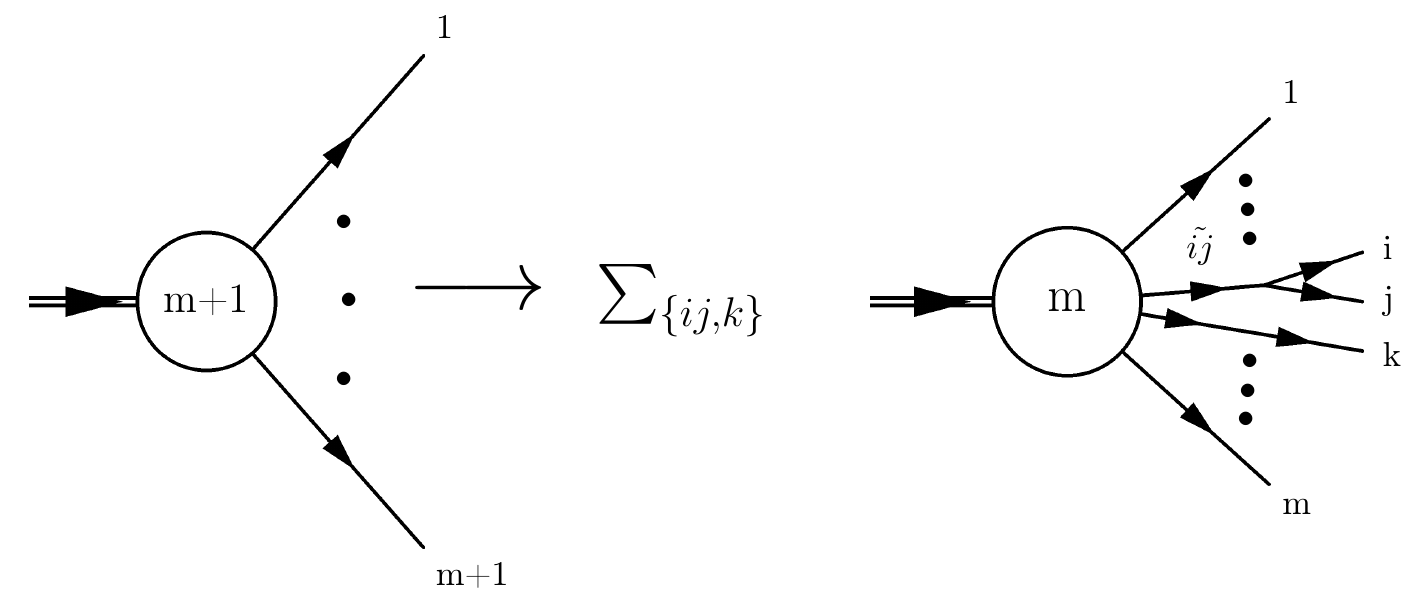}
	\caption{The dipole structure for a $2 \rightarrow m+1$ process.}
	\label{Dipole:Bild2}
\end{figure}

The general structure of the associated matrix element of $\mathrm{d}\sigma^{\text{A}}$ can then be rewritten as
\begin{eqnarray}
	\label{Dipole2}
	 \big| M^\text{A} \big|^2 &=& \sum_{i,j} \sum_{k\neq i,j} \mathcal{D}_{ijk} \\
		\nonumber &=&\sum_{i,j} \sum_{k\neq i,j} V_{ij,k}(p_i,\tilde{p}_{ij},\tilde{p}_{k})\otimes 		
		\big| M^{\text{B}}(\tilde{p}_{ij},\tilde{p}_{k}) \big|^2.
\end{eqnarray}
The universal product form on the right hand side mimics the factorization of $|M^{\text{R}}|^2$ in the soft and collinear limit. It encodes the two-step process of the Born-level production of an emitter-spectator pair with momenta $\tilde{p}_{ij}$ and $\tilde{p}_{k}$ followed by the decay of the emitter described by $V_{ij,k}$ as represented by the box in Fig.\ \ref{Dipole:Bild1}. 
The $V_{ij,k}$ are matrices in color and helicity-space of the emitter and the symbol $\otimes$ stands for phase space convolution and possible helicity and color sums between $V_{ij,k}$ and the exclusive Born-level matrix element 
$M^{\text{B}}(\tilde{p}_{ij},\tilde{p}_{k})$. They become proportional to the Altarelli-Parisi splitting functions in the collinear region and to eikonal factors in the soft region \cite{Dipole:Dipole2,Dipole:Splitting-functions}.

\begin{figure}[h]
	\includegraphics[scale=0.3]{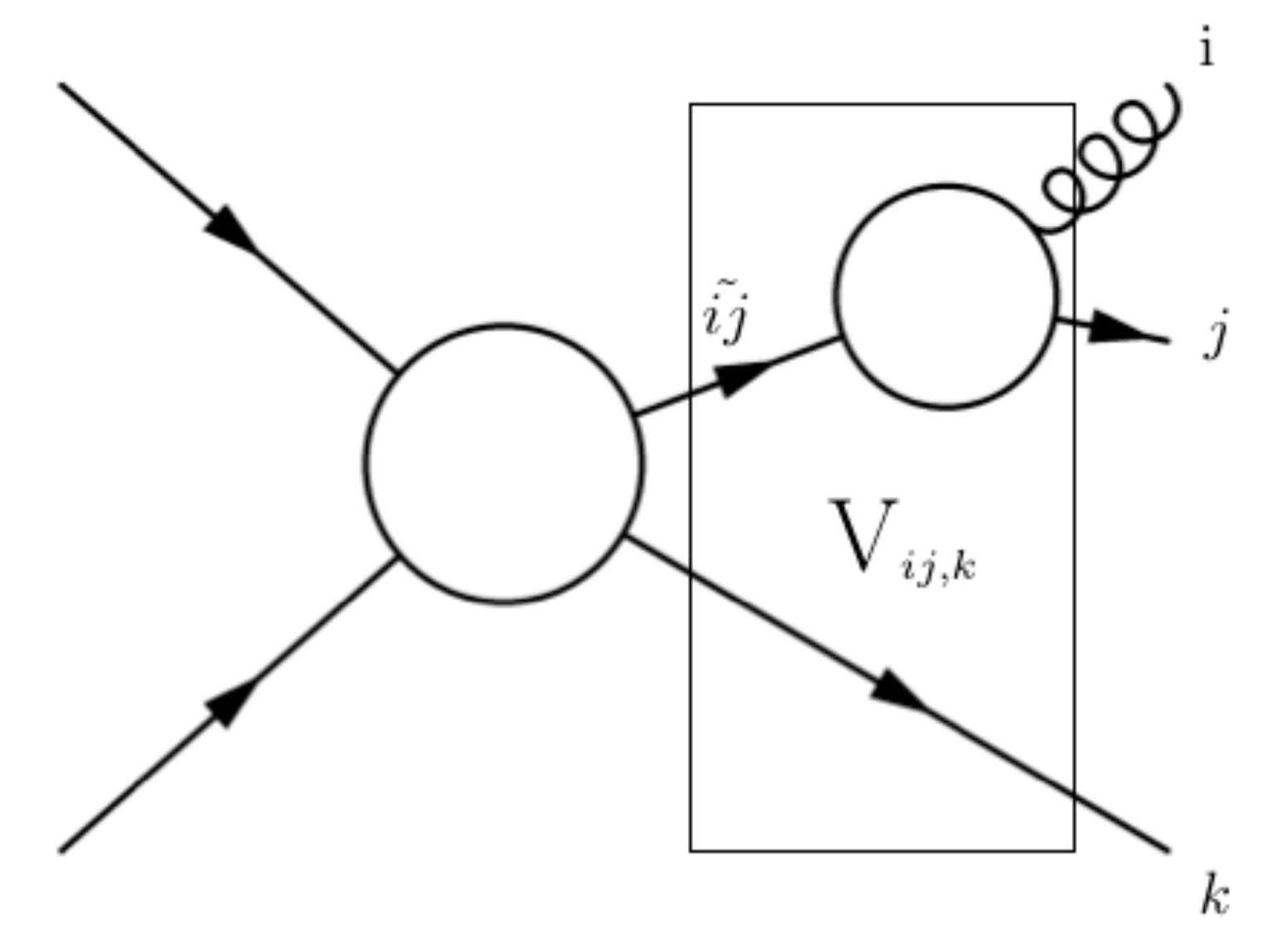}
	\caption{Factorization of a $2 \rightarrow 3$ process in the soft and collinear limit.}
	\label{Dipole:Bild1}
\end{figure}

In addition, Eq.\ (\ref{Dipole2}) allows for a factorizable mapping of the three-particle phase-space spanned by $p_i$, $p_j$, and $p_k$ onto the two-particle phase-space represented by the emitter and spectator momenta $\tilde{p}_{ij}(p_i,p_j,p_k)$ and $\tilde{p}_{k}(p_i,p_j,p_k)$ times the single gluon phase space spanned by $p_i$ as implied by Eq.\ (\ref{Dipole0}). It takes the form 
\begin{eqnarray}
	\label{Dipole3}
	\nonumber 		
	\tilde{p}_k^{\mu} & = &
	 	\frac{{\lambda^{1/2}\big( Q^2,m_{ij}^2,m_k^2\big)}}
			 {{\lambda^{1/2}\big( Q^2,(p_i+p_j)^2,m_k^2 \big)}} 
		\left( p_k^{\mu}-\frac{Q\cdot p_k}{Q^2}Q^{\mu} \right) \\
		& & + \frac{Q^2+m_k^2-m_{ij}^2}{2Q^2}Q^{\mu}
\end{eqnarray}
with
\begin{equation}
	\label{Dipole4} 
	\tilde{p}_{ij}^{\mu}=Q^{\mu}-\tilde{p}_k^{\mu}, \qquad
 	Q^{\mu}=p^{\mu}_i+p^{\mu}_j+p^{\mu}_k,
\end{equation}
and the emitter mass $m_{ij}$. For further definitions see App.\ \ref{Dipole:appendix}.
Since the $V_{ij,k}$ only describe the emitter decay and are essentially independent of the $2\rightarrow 2$ Born-level cross section part, they need to be calculated only once. In the case of (anti)quark-gluon splitting, the $V_{ij,k}$ are given by
\begin{widetext}
\begin{equation}
 	\label{Dipole6}
	\big\langle s \big| V_{ij,k} \big|s' \big\rangle ~=~ 
		2g_s^2 \mu^{2\epsilon}\ C_F \Biggr\{ \frac{2}{1-\tilde{z}_j(1-y_{ij,k})} - 	
		\frac{\tilde{v}_{ij,k}}{v_{ij,k}} \biggr[ 1 + \tilde{z}_j+\frac{m_q^2}{p_ip_j} + 		
		\epsilon \big( 1-\tilde{z}_j \big) \biggr] \Biggr\} \delta_{ss'}
		~=~ \big\langle V_{ij,k} \big\rangle \delta_{ss'}
\end{equation}
with $i=g$, $j=q$ (or $\bar{q}$) and $k=\bar{q}$ (or $q$). $s$ and $s'$ are the emitter-spins, $g_s$ is the strong coupling, and $C_F=4/3$ for $SU(3)_{\text{Color}}$. The dimensional regularization scale $\mu$,
which drops out of the final result $\sigma_{\text{NLO}}$, is set to be equal to the renormalization scale.
For further definitions see App.\ \ref{Dipole:appendix}. The auxiliary cross section $\mathrm{d}\sigma^{\text{A}}$ can then be constructed using Eq.\ (\ref{Dipole6}) together with Eq.\ (\ref{Dipole2}).

The virtual dipole contributions of Eq.\ (\ref{Dipole0}) can be rewritten as 
\begin{equation}
 	\label{Dipole7}
 	\int_2 \Big[ \mathrm{d}\sigma^{\text{V}} + \int_1 \mathrm{d}\sigma^{\text{A}} \Big]_{\epsilon=0} = 
	\int_2 \Big[ \mathrm{d}\sigma^{\text{V}} + \mathrm{d}\sigma^B \otimes \mathbf{I}_2 
		\left(\epsilon,\mu^2;\{p_a,m_a\}\right) \Big]_{\epsilon=0},
 \end{equation}
where the index $a$ runs over the two Born-level final states.
Following Eq.\ (\ref{Dipole2}), due to the phase-space mapping denoted in Eqs.\ (\ref{Dipole3}) and (\ref{Dipole4}), ${\rm d}\sigma^{\mathrm{A}}$ explicitly depends on the gluon phase-space spanned by $p_i$ only via the universal factor $V_{ij,k}$. Therefore this integration can be performed analytically once and for all as given in Eq.\ (\ref{Dipole7}). Following again Ref.\ \cite{Dipole:Dipole1} for final state radiation, the function $\mathbf{I}_2$ can then be written as 
\begin{eqnarray}
 	\label{Dipole9}
	\nonumber 		
	\mathbf{I}_2 \left( \epsilon,\mu^2; \{p_a,m_a\} \right) & = & 
		-\frac{g_s^2}{8\pi^2}\frac{(4\pi)^{\epsilon}}{\mathbf{\Gamma}(1-\epsilon)}\sum_j\frac{1}{\mathbf{T}_j^2}
		\sum_{k\neq j}\mathbf{T}_j\cdot \mathbf{T}_k \, 
		\Biggr[ \mathbf{T}_j^2 \, \biggr( \frac{\mu^2}{s_{jk}} \biggr)^{\!\!\epsilon} \, \biggr( \mathcal{V}_j \left( s_{jk},m_j,m_k;\epsilon \right) - \frac{\pi^2}{3} \biggr) \\  
	& & + \mathbf{\Gamma}_j(\mu,m_j;\epsilon) 
+\gamma_j \,\log \! \left( \frac{\mu^2}{s_{jk}} \right) + \gamma_j+\text{K}_j+O(\epsilon)\Biggr] ,
\end{eqnarray}
\end{widetext}
where $s_{jk}=2p_jp_k$, the $\mathbf{T}$ are the color charges in the representation of the associated particle, and $j$ and $k$ run over all possible emitter-spectator combinations as in Eq.\ (\ref{Dipole1}). For the expressions of $\mathbf{\Gamma}_j$, $\gamma_j$ and $\text{K}_j$ see App.\ \ref{Dipole:appendix}.
The function $\mathcal{V}_j$ can be further decomposed into a ($j \leftrightarrow k$)-symmetric and singular (S) and a non-symmetric and non-singular (NS) part
\begin{eqnarray}
	\label{Dipole10}
	\nonumber \mathcal{V}_j(s_{jk},m_j,m_k;\epsilon) & = & 
		\mathcal{V}^{(\text{S})}_j(s_{jk},m_j,m_k;\epsilon) \\
		& & + \mathcal{V}^{(\text{NS})}_j(s_{jk},m_j,m_k). 
		~~~~~~~~
\end{eqnarray}
Dealing with light quarks, numerical instabilities of the relevant dipoles are encountered if the mass of a final state quark drops far below the initial state energy, e.g., for first generation quarks. Therefore we take the light quarks ($u$, $d$ and $s$) as massless. That makes it necessary to also take into account the associated $\mathcal{V}^{(\text{S})}$ and $\mathcal{V}^{(\text{NS})}$, which are given in  App.\ \ref{Dipole:appendix} for all possible combination of emitter-spectator masses.

These are all the parts needed for calculating the relevant dipole contributions. The concrete calculation of the various elements has been performed in a semi-automatic way using \FeynCalc\ \cite{FeynCalc} for simplifying the Dirac algebra. Furthermore, since we are dealing with initial-state neutralinos which are Majorana fermions, one needs to take care of handling the fermion flow correctly. This has been done following Ref.\ \cite{Dipole:Majorana}.

In addition, we have introduced a minimal width for squark propagators to avoid numerical instabilities due to quasi-collinear singularities in the $t$- and $u$-channels. This minimal value is set to 10 GeV. For our typical scenarios we have explicitly checked that on the one hand this renders the three-particle phase-space integration stable, but on the other hand has no relevant impact on the relic density.

%% file: results.tex
\section{Numerical results}
\label{Numerics}

\subsection{Impact on the cross section}
\label{CrossSectionTeil}

\begin{figure*}
	\includegraphics[width=0.49\textwidth]{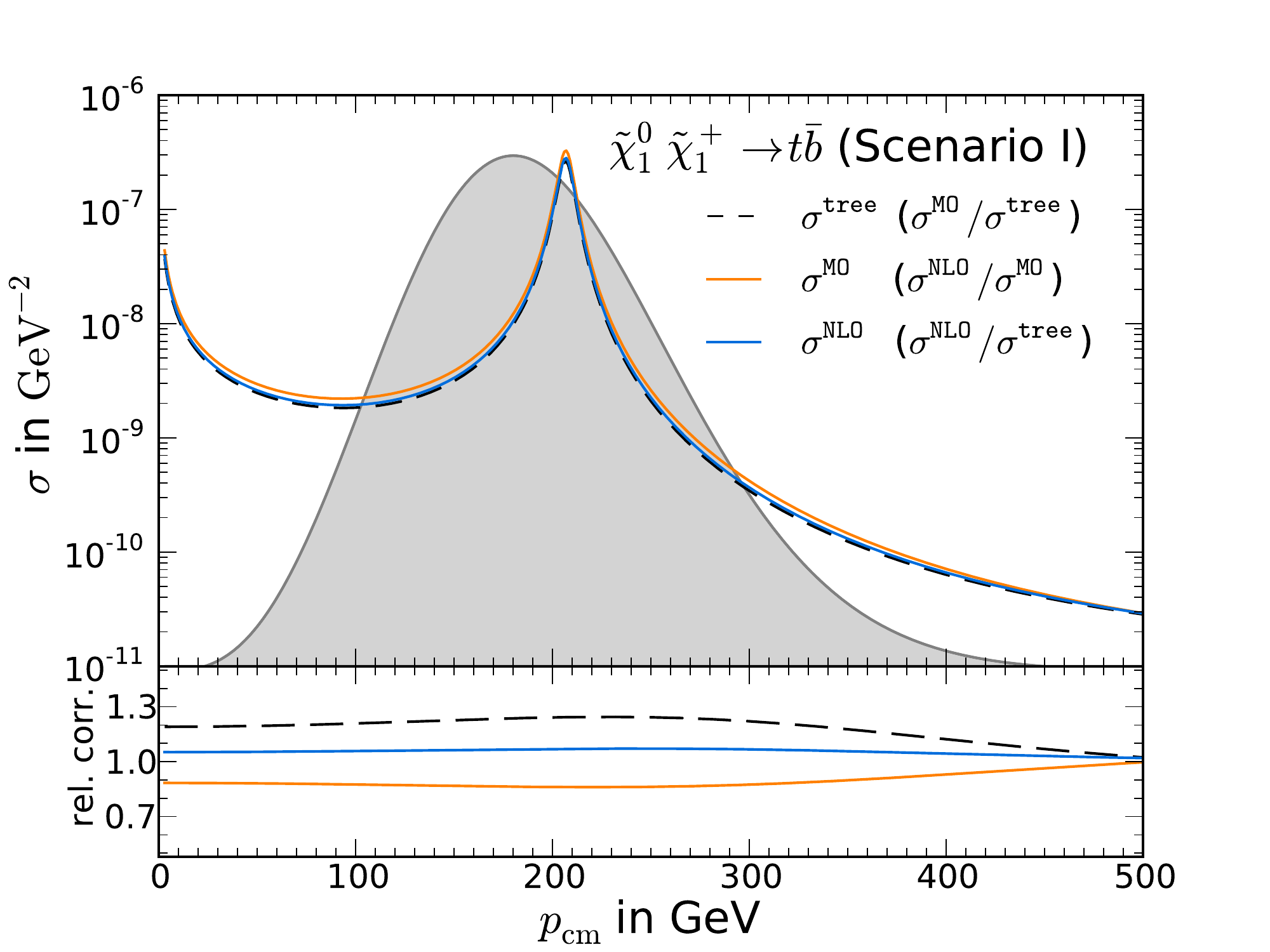}
	\includegraphics[width=0.49\textwidth]{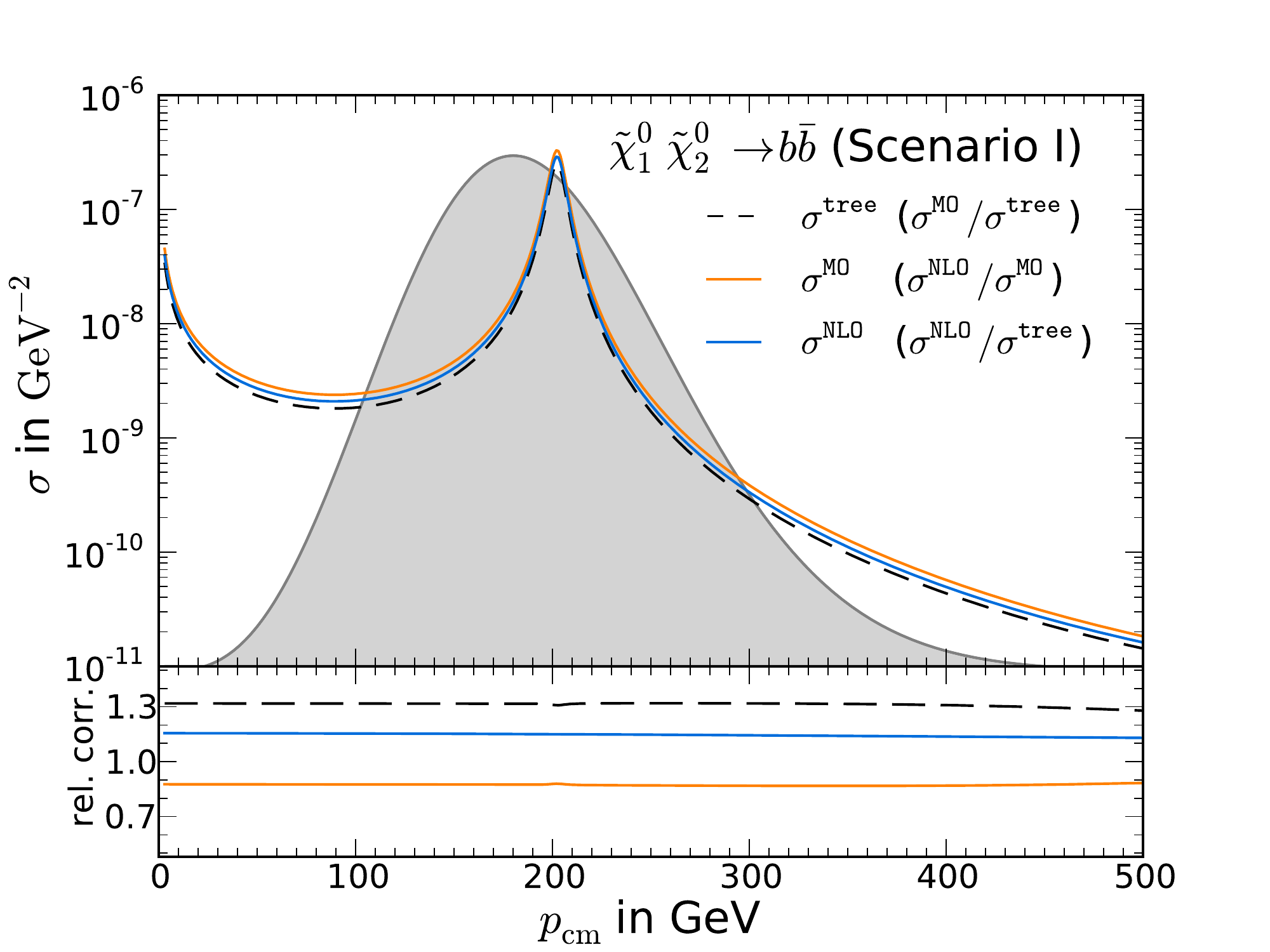}
	\includegraphics[width=0.49\textwidth]{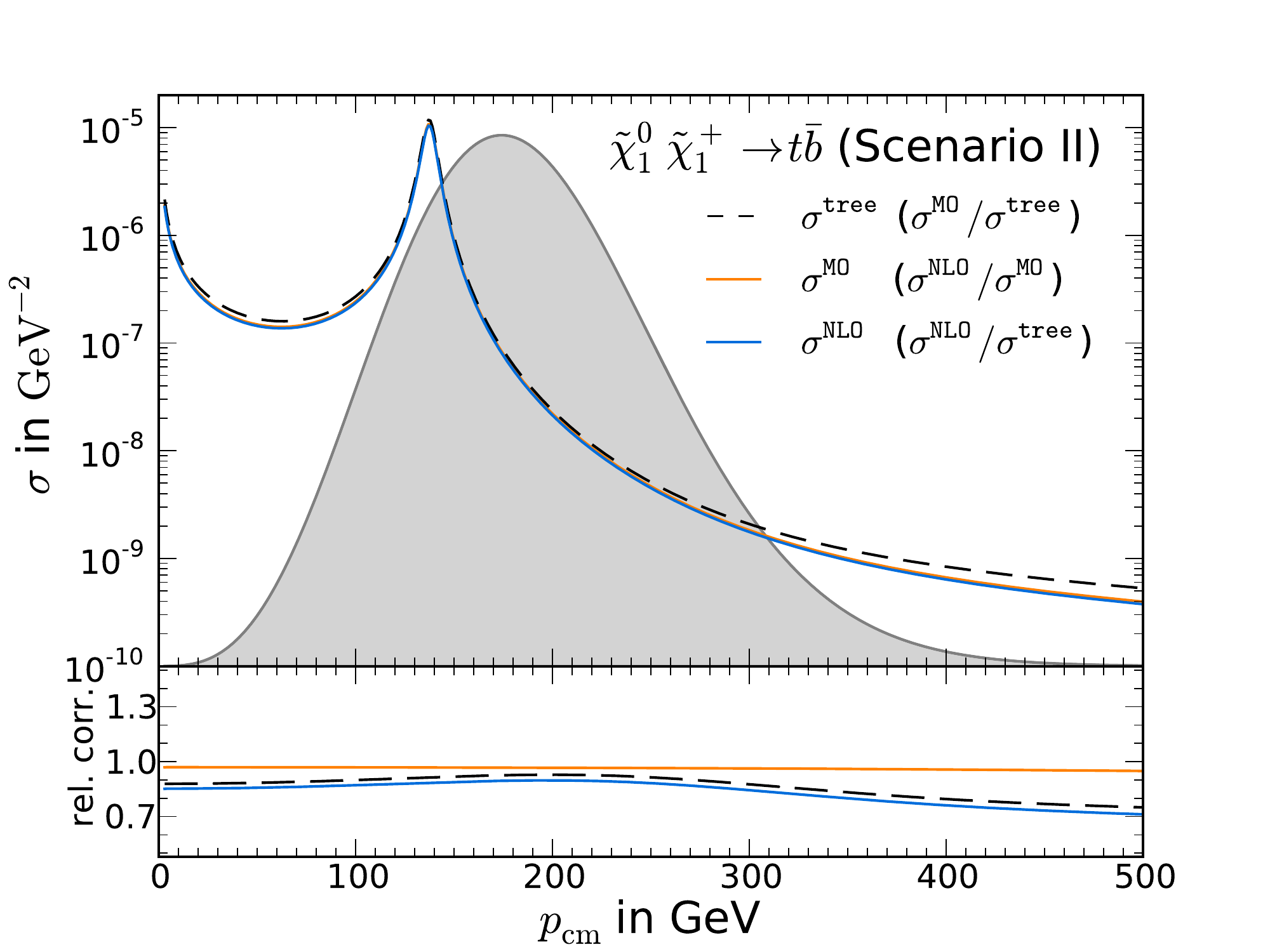}
	\includegraphics[width=0.49\textwidth]{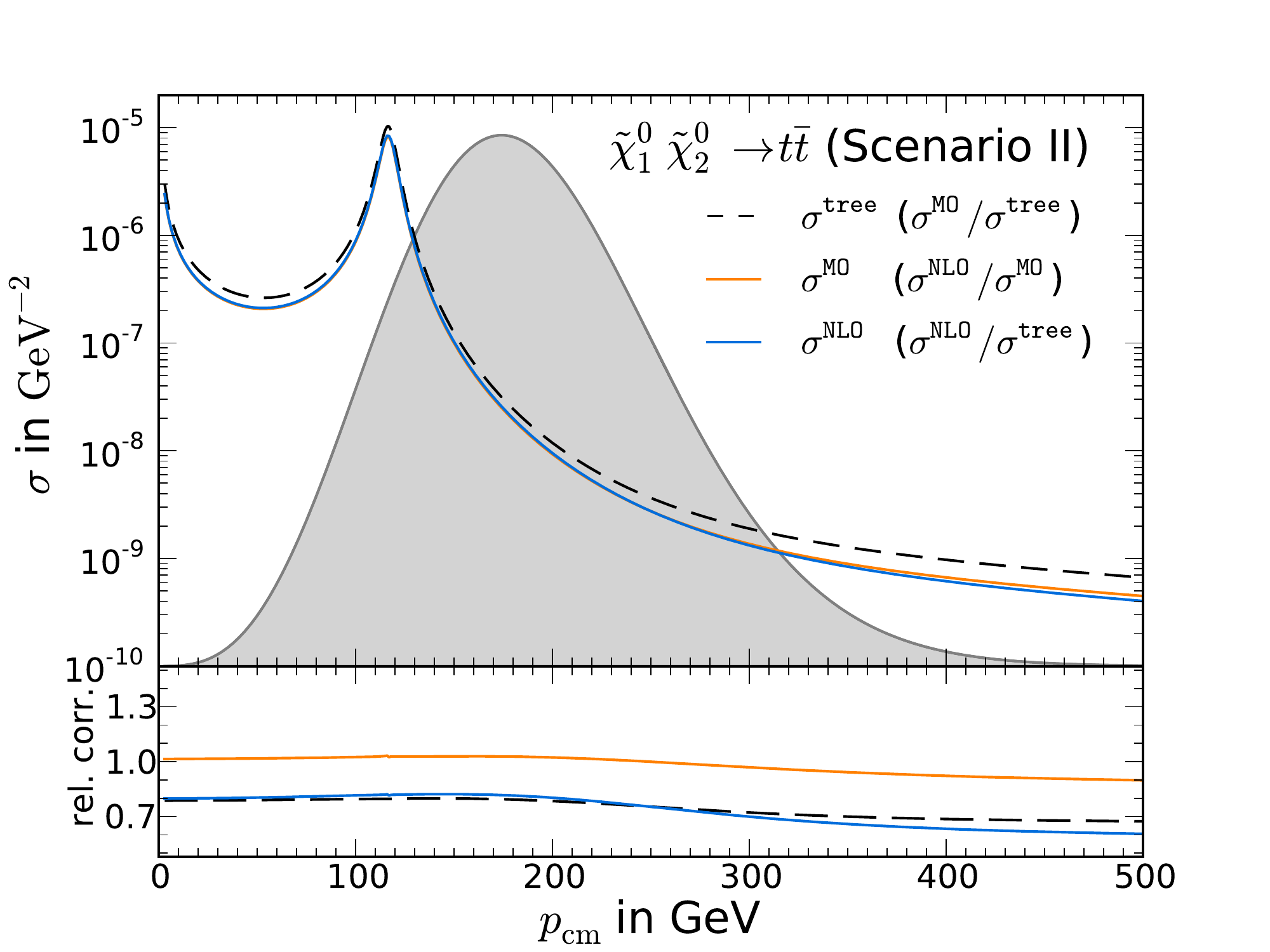}
	\includegraphics[width=0.49\textwidth]{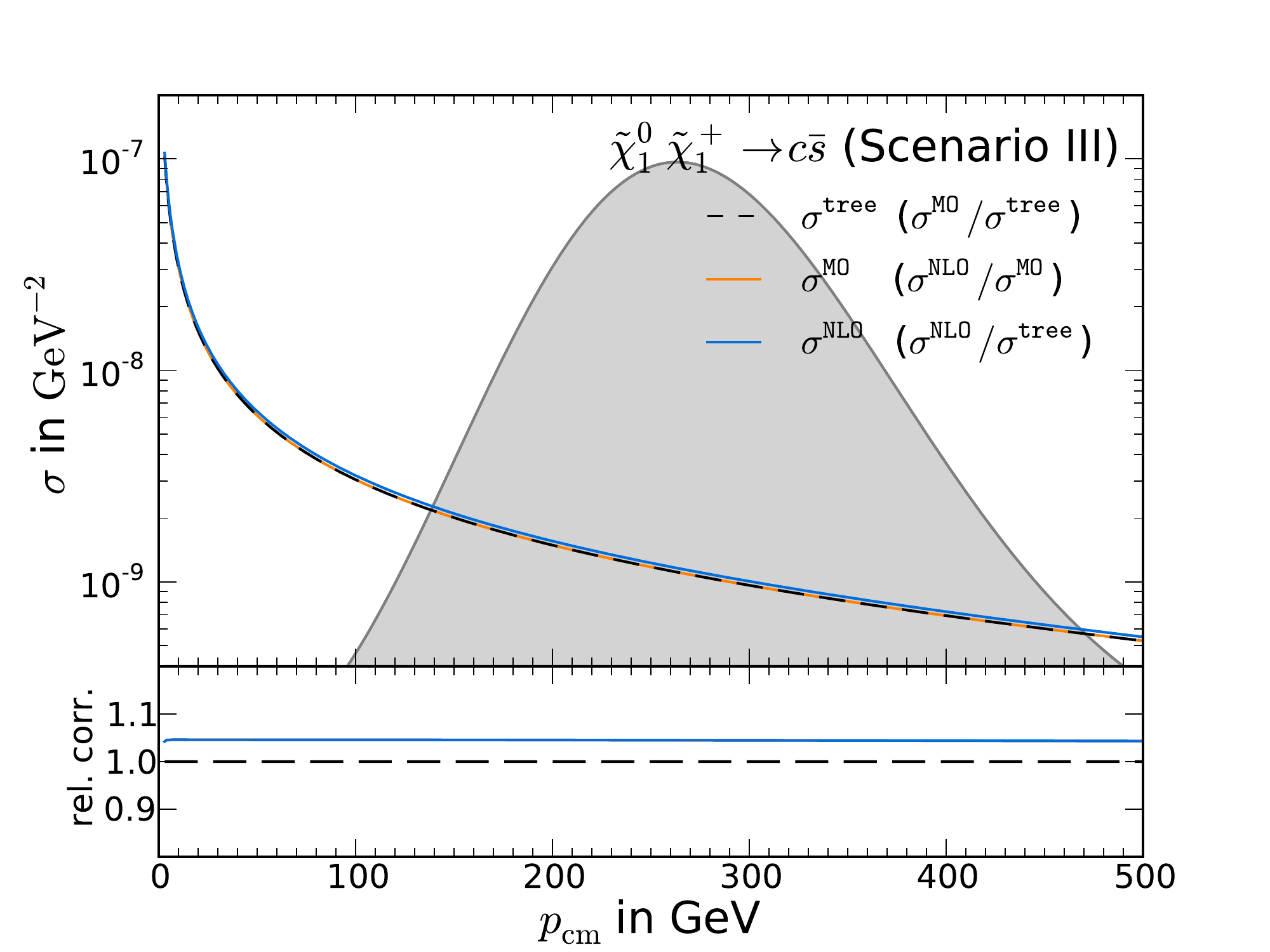}
	\includegraphics[width=0.49\textwidth]{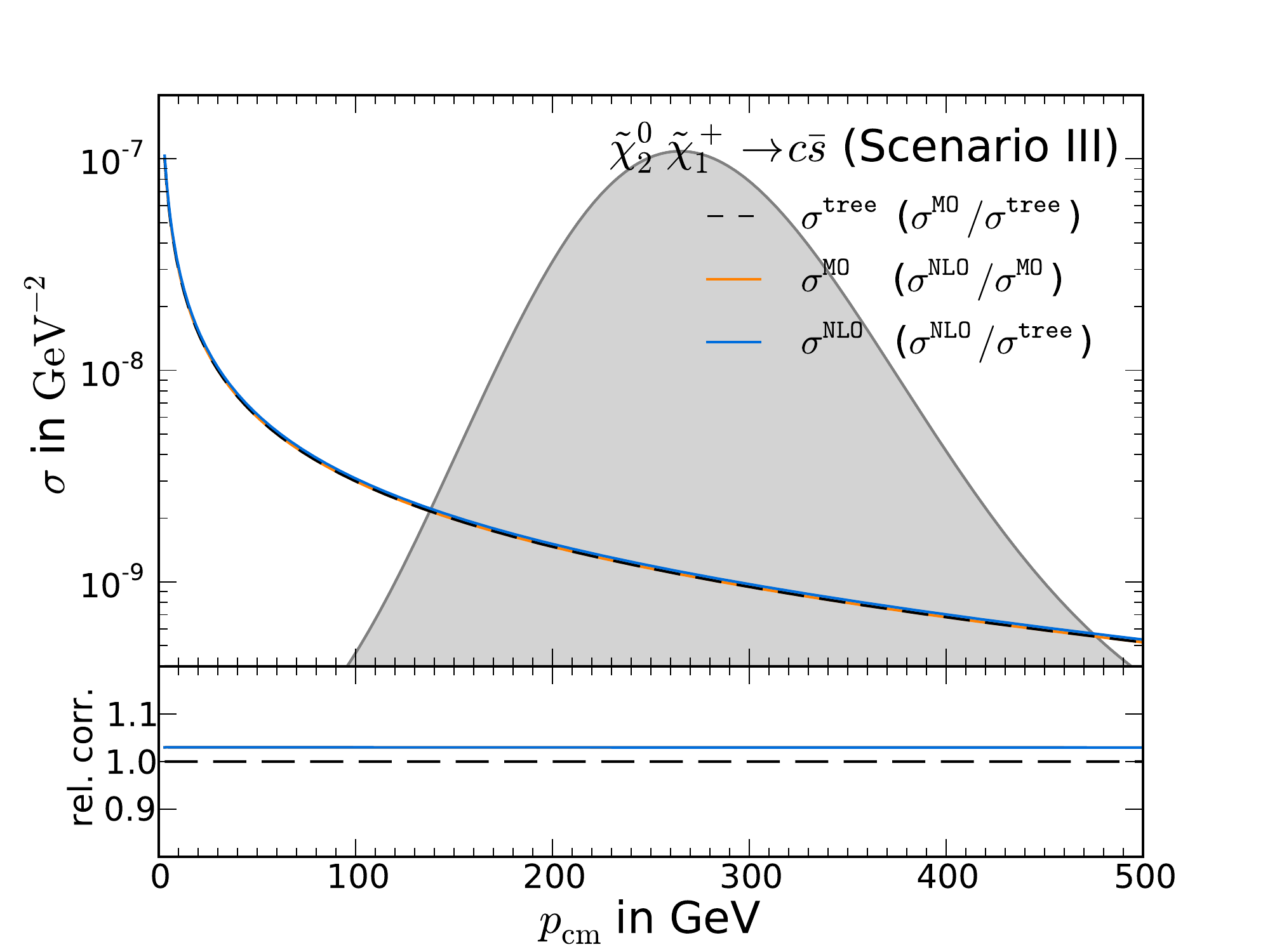}
	\caption{Tree level (black dashed line), full one-loop (blue solid line) and \MO\ (orange solid line) cross sections for selected channels in the scenarios of Tab. \ref{ScenarioList}. The upper part of each plot shows the absolute value of $\sigma$ in GeV$^{-2}$ in dependence of the momentum in the center-of-mass frame $p_{\mathrm{cm}}$. The gray areas indicate the thermal distribution (in arbitrary units). The lower parts of the plots show the corresponding ratios of the cross sections (second item in the legends).}
	\label{fig:CrossSectionPlots}
\end{figure*}

In this Section we discuss the impact of our full $\mathcal{O}(\alpha_s)$ corrections on the cross sections of the processes in Eqs.\ (\ref{NeuNeuAnni}) -- (\ref{ChaChaAnni}). In Fig.\ \ref{fig:CrossSectionPlots} we present the cross sections of the most relevant gaugino (co-)annihilation channels of Tab.\ \ref{ScenarioChannels} against $p_{\mathrm{cm}}$, the momentum in the center-of-mass frame. We show the cross sections at tree-level (black dashed line), at one loop (blue solid line), and the corresponding value obtained with \MO/{\tt CalcHEP} (orange solid line). 
Also shown as gray shaded areas are the Boltzmann velocity distributions
of the dark matter particles (in arbitrary units). Their maxima
may coincide with the maximal cross sections (induced, e.g., by Higgs
resonances as in the upper two plots), but depending on the particle
masses the maximal cross section often also sits on the shoulder of
the Boltzmann distribution (cf.\ the two central plots).
The lower parts of each plot show the different ratios between the three cross sections (second item in the legends).

The upper left plot of Fig.\ \ref{fig:CrossSectionPlots} shows the cross sections for the channel $\tilde{\chi}^0_1\tilde{\chi}^+_1\rightarrow t \bar{b}$ for scenario I. The most striking feature is the large $H^+$-resonance at $p_{\mathrm{cm}} \sim 200$ GeV, which we have already discussed in Sec.\ \ref{Pheno}. The curves of our tree level and one-loop cross sections lie rather closely together, they differ by 5--10\% as can be seen in the lower part of the graph. However, the deviation between our tree level and the \MO\ calculation accounts for more than 20\% (black dashed line in the lower plot). This is mainly due to the different treatment of the third-generation quark masses as discussed in Sec.\ \ref{Technical}.

In the upper right corner, we show the analogous plot for the channel $\tilde{\chi}^0_1\tilde{\chi}^0_2\rightarrow b \bar{b}$. As before, we observe a rather large resonance at $p_{\mathrm{cm}}\sim$ 200 GeV, which is this time due to the $s$-channel co-annihilation via $A^0$. In comparison to the previous plot, the three cross sections have larger deviations from each other. Our loop corrections increase the tree level result by almost 20\%, while the \MO\ cross section differs by about 35\% from our tree level. Again, this is due to the different treatment of the bottom quark. Note that the inclusion of the NLO corrections shifts the cross section towards the effective tree level of \MO.

The two center plots of Fig.\ \ref{fig:CrossSectionPlots} show the cross sections for the channels $\tilde{\chi}^0_1\tilde{\chi}^+_1\rightarrow t \bar{b}$ and $\tilde{\chi}^0_1\tilde{\chi}^0_2\rightarrow t \bar{t}$ for scenario II. Once more we find large resonances at $p_{\mathrm{cm}}\sim$ 140 GeV and $p_{\mathrm{cm}}\sim$ 120 GeV, which are due to $H^+$ and $H^0$ exchange in the $s$-channel. The positions of the peaks differ more than in scenario I. This is due to the fact that the masses $m_{\tilde{\chi}^+_1}$ and $m_{\tilde{\chi}^0_2}$ as well as $m_{H^0}$ and $m_{H^+}$, respectively, are less degenerate than in scenario I (see Tab.\ \ref{ScenarioProps} and the discussion in Sec.\ \ref{Pheno}). Here, our one-loop corrections lead to a change in the cross section by 10--30\% for the channel $\tilde{\chi}^0_1\tilde{\chi}^+_1\rightarrow t \bar{b}$ and by remarkable 20--40\% for the channel $\tilde{\chi}^0_1\tilde{\chi}^0_2\rightarrow t \bar{t}$. However, note that these corrections are in this case already well approximated by the efficient treatment of the effective quark masses in the \MO\ tree level calculation. The blue and orange lines overlap in the upper parts and the black dashed line basically follows the blue line in the lower parts of the plots.

The remaining two plots depict the cross sections for the channels $\tilde{\chi}^0_1\tilde{\chi}^+_1\rightarrow c \bar{s}$ and $\tilde{\chi}^0_2\tilde{\chi}^+_1\rightarrow c \bar{s}$ for scenario III. Both of the diagrams are very similar. In the absence of resonances, the cross sections drop monotonously with increasing center-of-mass energy. Nevertheless, it is important to note that in these cases our tree level perfectly agrees with the \MO\ cross section. This can easily be seen in the lower parts of the plots, where the black dashed lines correspond to the constant value one. As our renormalization scheme does not modify the masses of the $c$ and $s$ quarks, this matches our expectations. The loop corrections can be read off from the blue lines (which completely overlaps with the orange lines) in the lower plots and account for roughly 5\% and 3\% respectively. They are rather small which is due to the fact that the involved SUSY particles are relatively heavy in this scenario (see Tab.\ \ref{ScenarioList}).

We close this Subsection with a remark on the treatment of the particle widths in resonant propagators. In our code, the widths are always active, whereas in \MO/{\tt CalcHEP} they are switched on only in a rather narrow interval around the resonance. In order to compare our calculation to the one implemented in \MO, we have modified the treatment of the width in {\tt CalcHEP} such that it is taken into account over the full range of $p_{\rm cm}$ shown in Fig.\ \ref{fig:CrossSectionPlots}. For the following calculation of the relic density, however, we have not modified the treatment of the width in CalcHEP.

\subsection{Impact on the relic density}

As already mentioned, we have calculated the full $\mathcal{O}(\alpha_s)$ corrections to all processes in Eqs.\ (\ref{NeuNeuAnni}) -- (\ref{ChaChaAnni}). Note that, e.g., Eq.\ (\ref{NeuNeuAnni}) alone covers ten possible initial states combined with six independent final states, i.e.\ 60 individual channels in total. All of these channels contribute to the annihilation cross section of Eq.\ (\ref{sigann}), but most of them are marginal. Unfortunately the important channels are not known a priori and depend strongly on the individual scenario. The NLO calculations are much more CPU-intensive than the corresponding tree-level calculations, especially the integration of real radiation over the three-particle phase space. We have therefore constructed a filtering mechanism allowing to perform efficient numerical analyses on the neutralino relic density. If the tree-level contribution of a single process in Eqs.\ (\ref{NeuNeuAnni}) -- (\ref{ChaChaAnni}) to the annihilation cross section is at least 2\%, we exchange the corresponding \CHep\ cross section by our result at NLO. The less relevant processes of Eqs.\ (\ref{NeuNeuAnni}) -- (\ref{ChaChaAnni}) are replaced by ours at tree level for consistency. All other processes, e.g., those including electroweak final states, are unchanged.

To investigate the impact of the loop corrections on the neutralino relic density, we have scanned the $M_1$--$M_2$ and $M_1$--$\mu$ planes surrounding the reference scenarios I and II using the filter mechanism described above and keeping all other pMSSM parameters fixed to the values of Tab.\ \ref{ScenarioList}. As a prelude to this discussion, we analyse which gaugino annihilation and coannihilation channels are dominant in these planes. 

\begin{figure*}
	\includegraphics[width=0.49\textwidth]{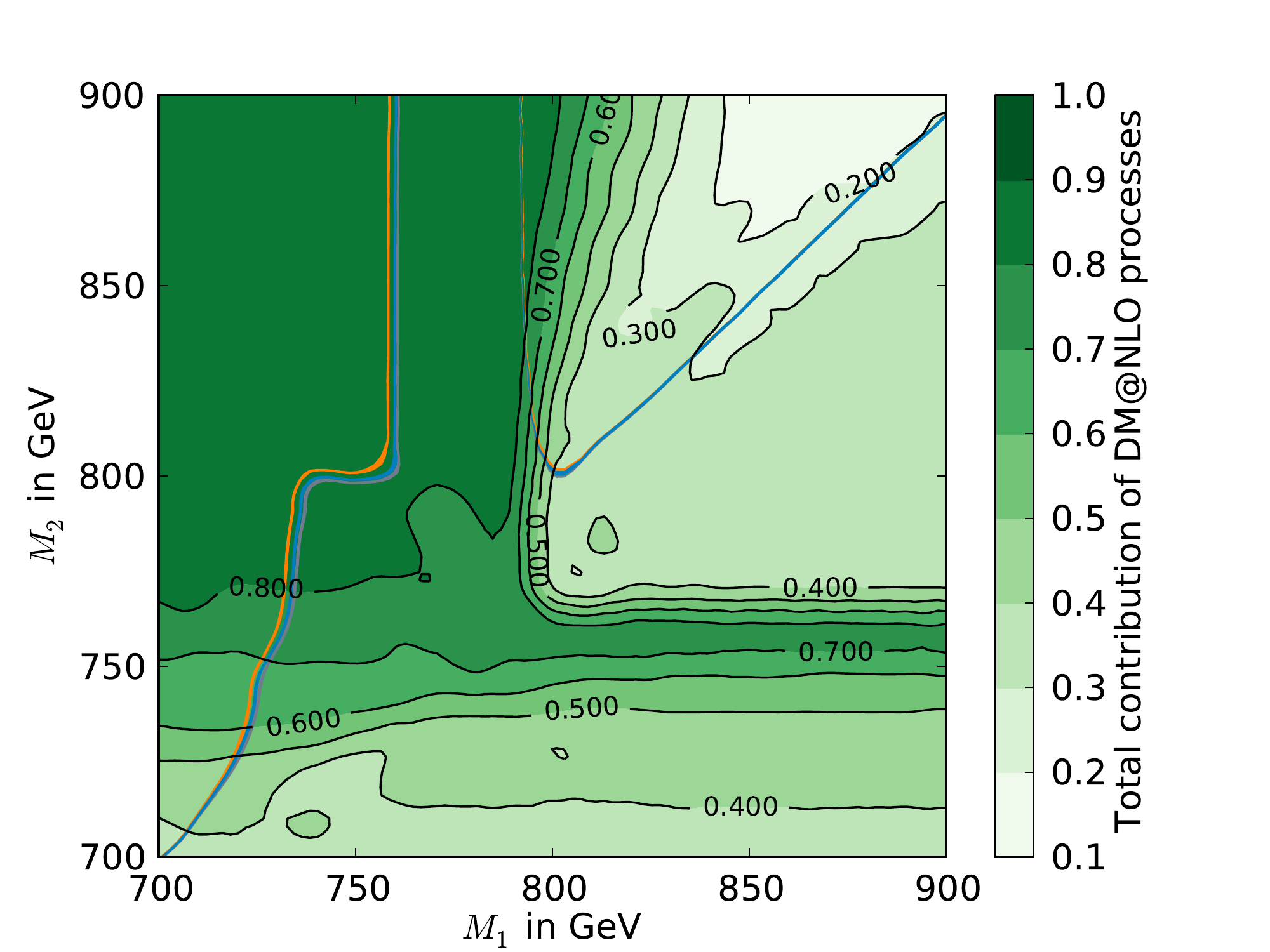}
	\includegraphics[width=0.49\textwidth]{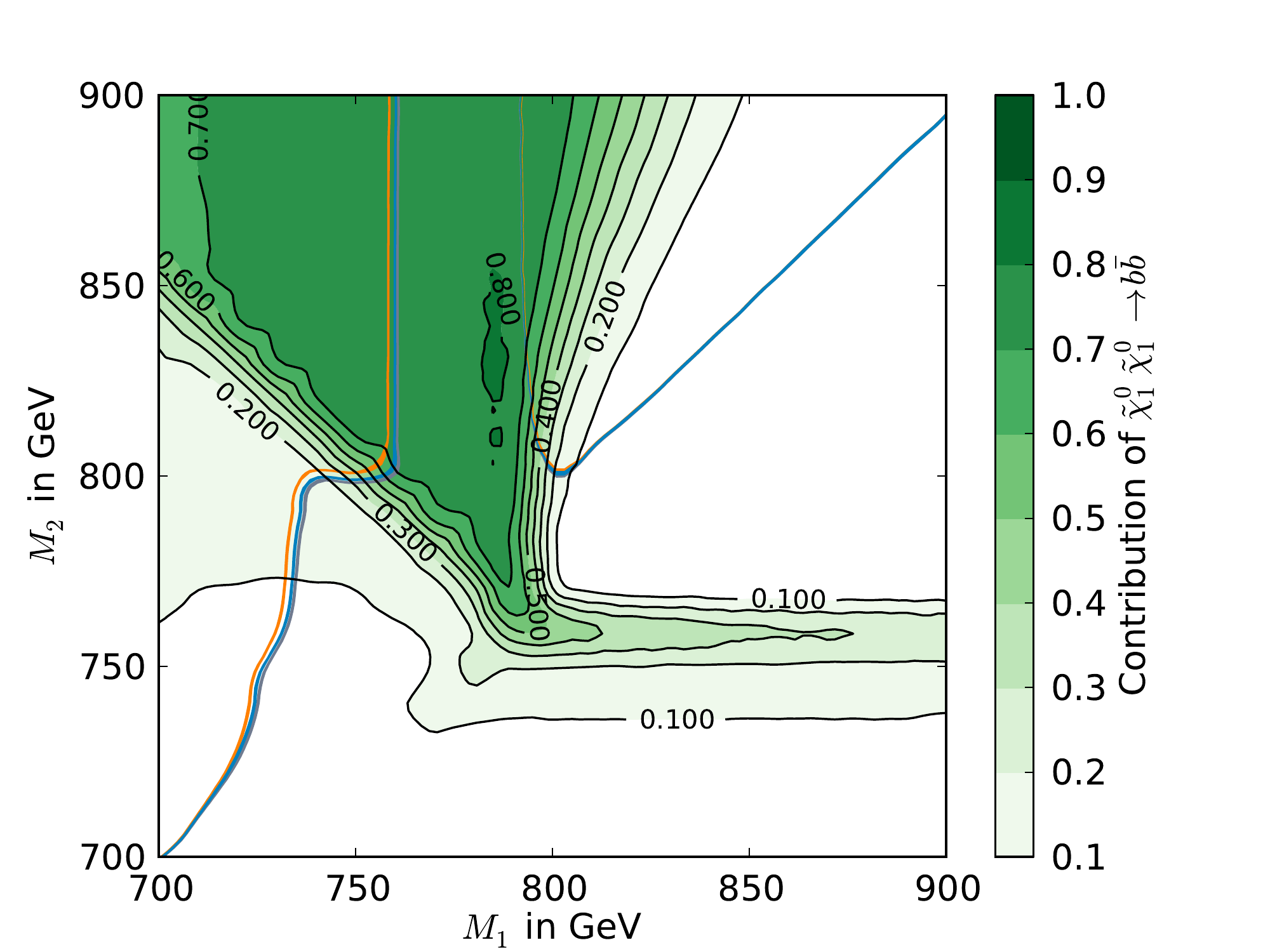}
	\includegraphics[width=0.49\textwidth]{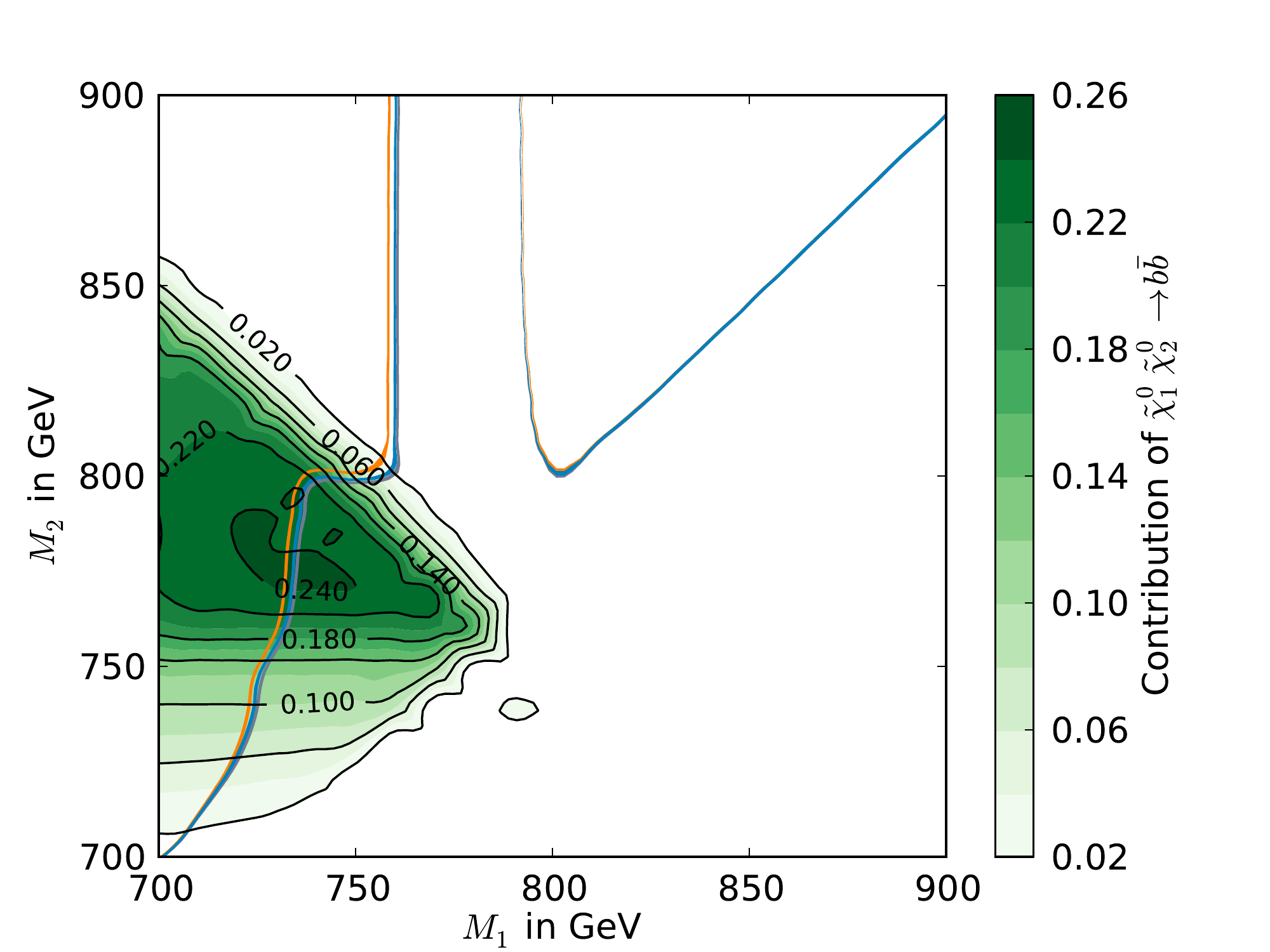}
	\includegraphics[width=0.49\textwidth]{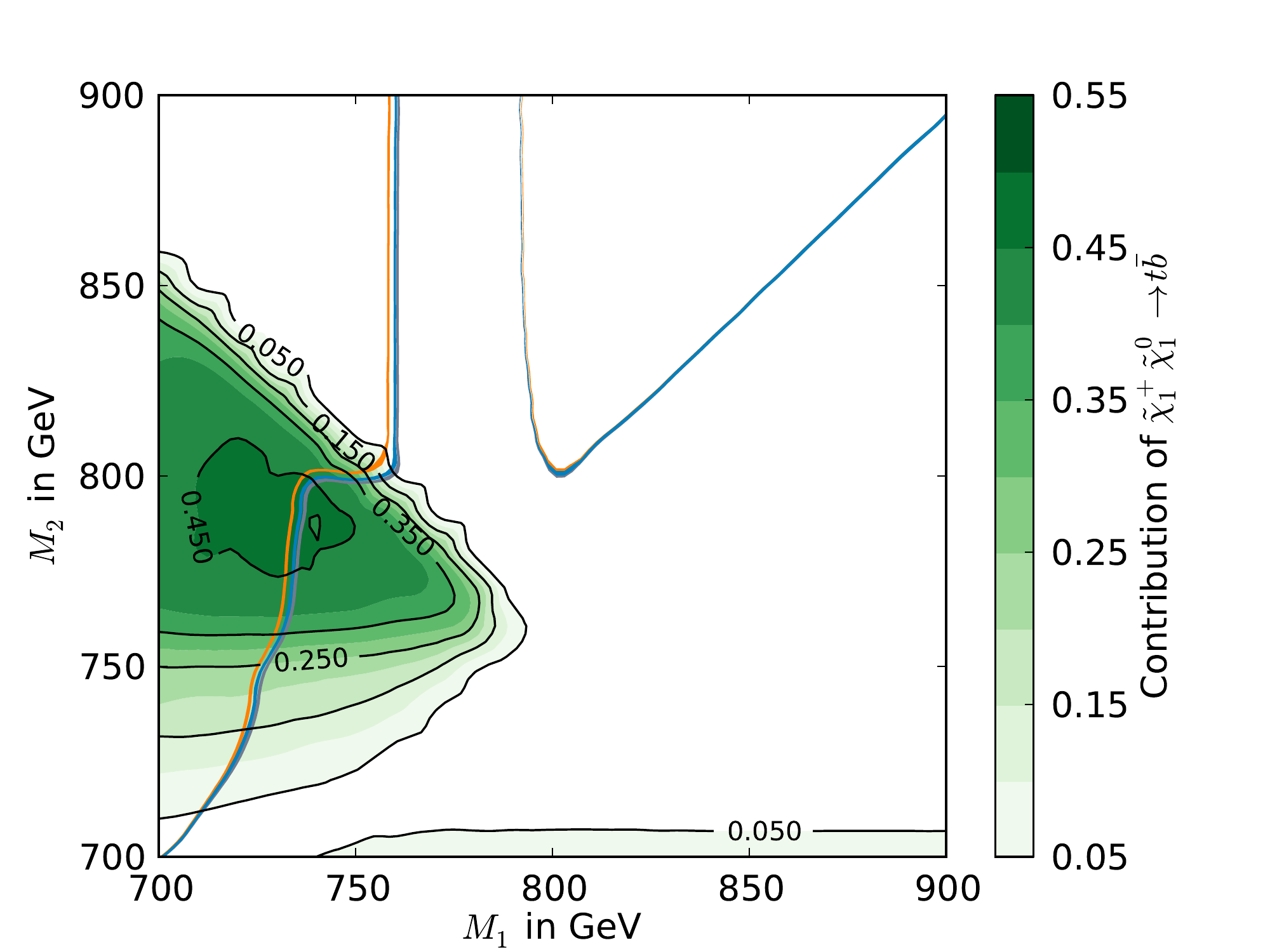}
	\includegraphics[width=0.49\textwidth]{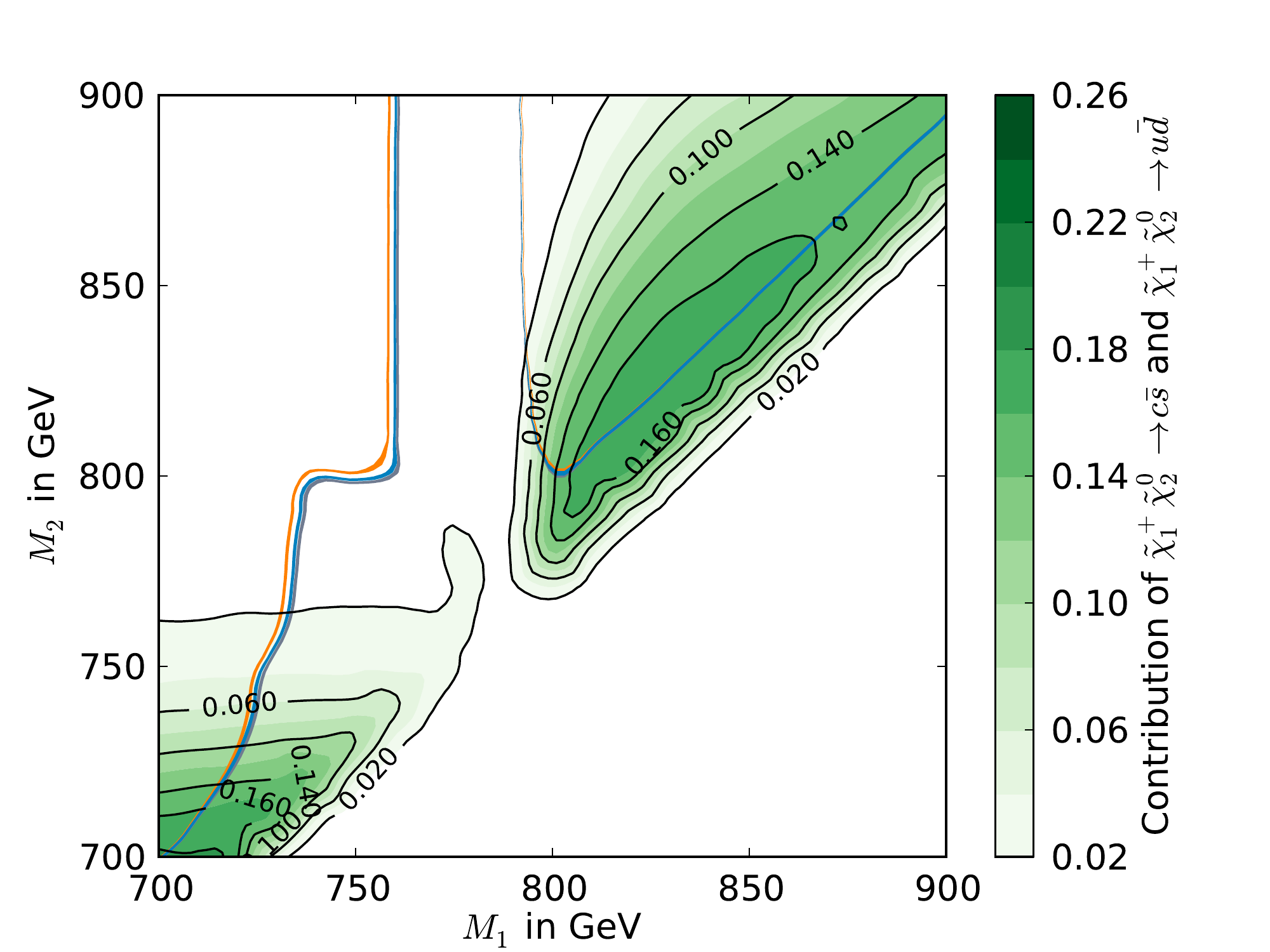}
	\includegraphics[width=0.49\textwidth]{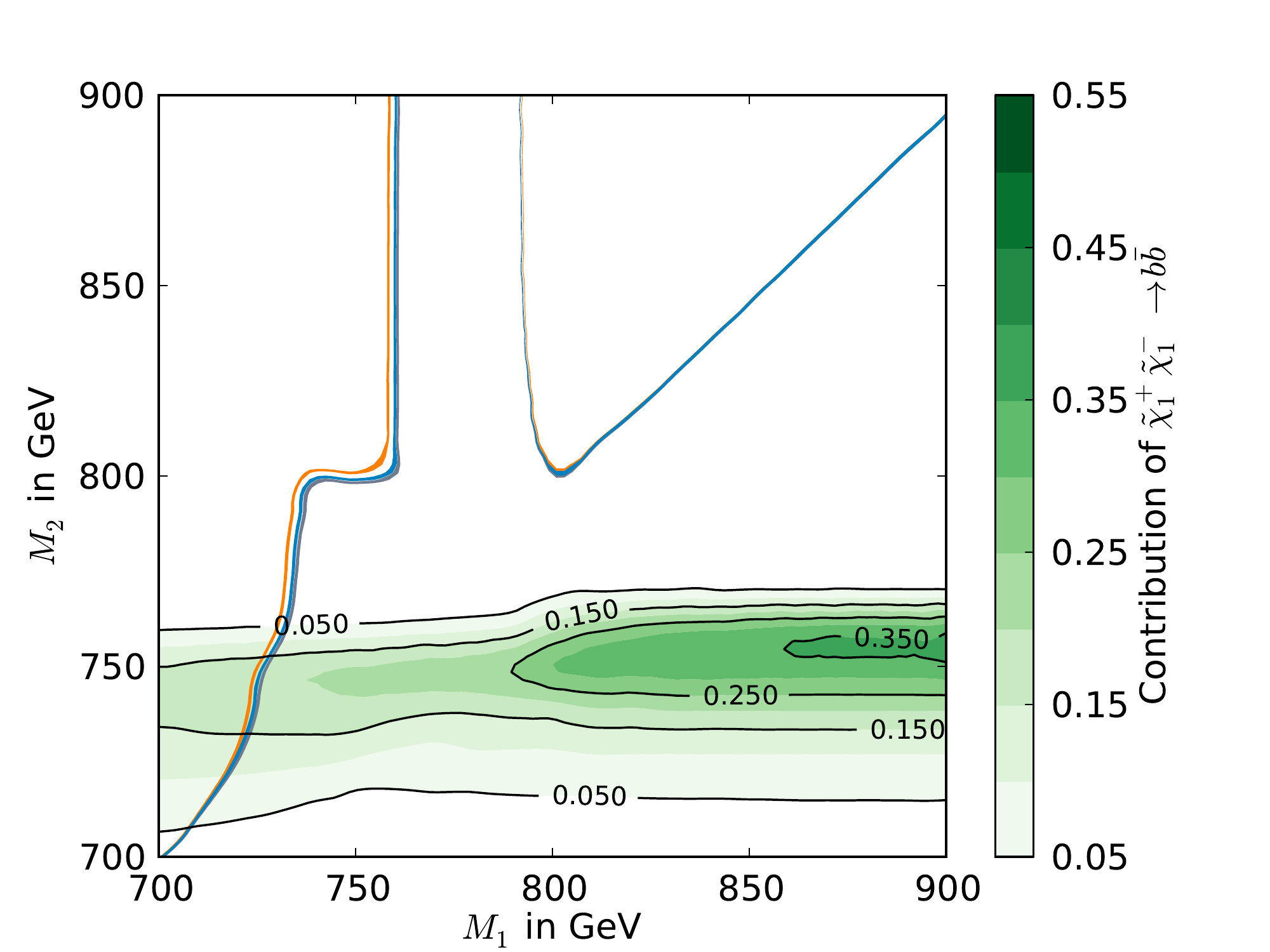}
	\caption{Relative importance of the processes that contribute to the neutralino relic density in the $M_1$--$M_2$ plane surrounding scenario I. The three colored lines represent the part of the parameter space which leads to a neutralino relic density compatible with the Planck limits given in Eq.\ (\ref{Planck}) using the standard \MO\ calculation (orange), our tree-level calculation (grey), and our full one-loop calculation (blue).}
	\label{fig:G5Plots}
\end{figure*}

In the upper-left plot of Fig.\ \ref{fig:G5Plots}, for scenario I, we show in green the total contribution of the processes that we improve to NLO. This contribution reaches more than 80\% in the upper-left corner and drops to less than 20\% in the upper right corner, where electroweak final states become dominant. The three colored lines represent the part of the parameter space which leads to a neutralino relic density compatible with the Planck limits given in Eq.\ (\ref{Planck}). The orange line indicates the standard \MO\ result, the grey one corresponds to our tree level calculation, and the blue one represents our full one-loop calculation. The thinness of these curves may be seen as a visualization of how constraining this limit actually is, if one assumes the neutralino to fully explain the CDM. We will focus on the analysis of the annihilation and coannihilation channels first and return to the discussion of the relic density later.

\begin{figure*}
	\includegraphics[width=0.49\textwidth]{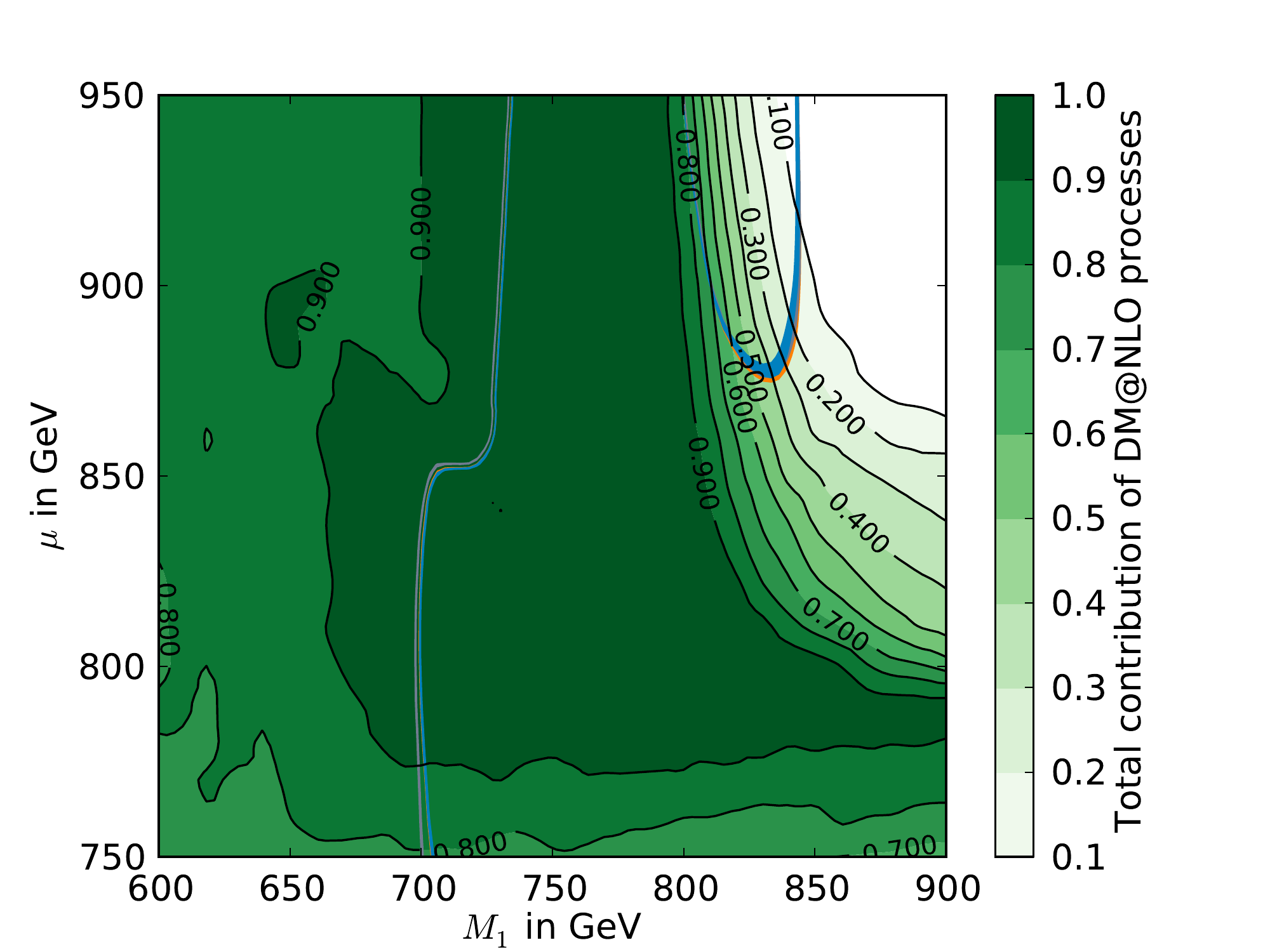}
	\includegraphics[width=0.49\textwidth]{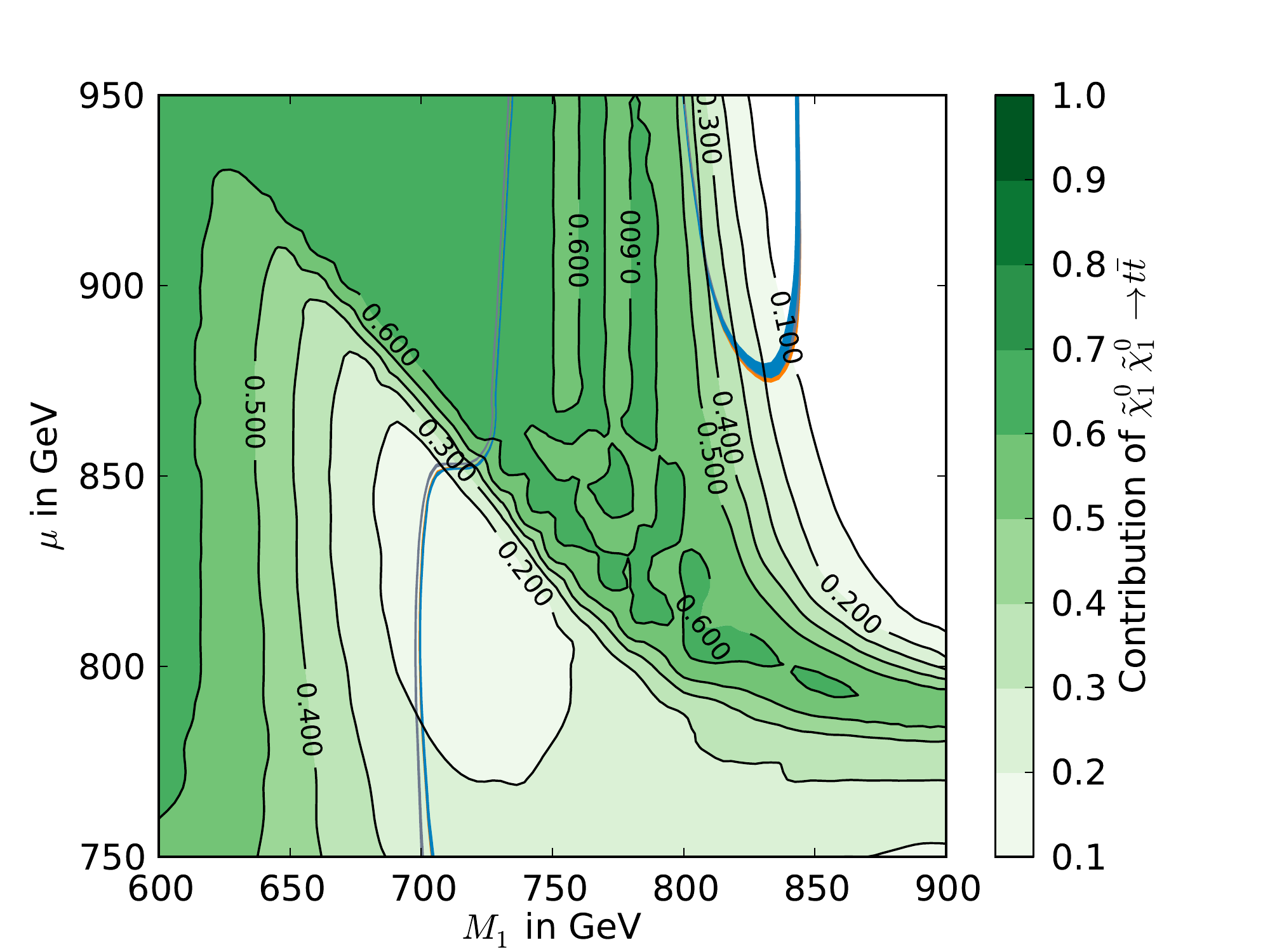}
	\includegraphics[width=0.49\textwidth]{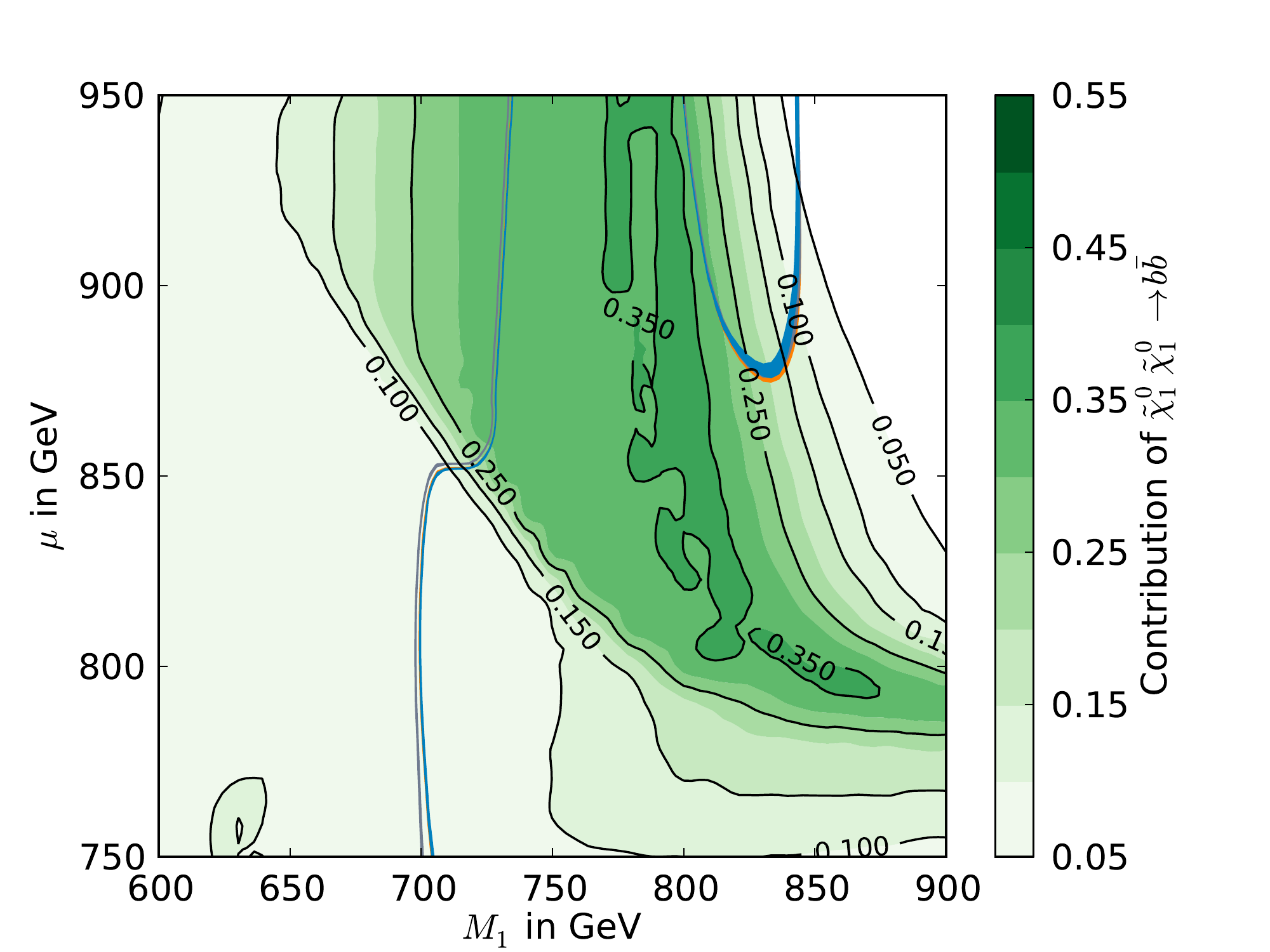}
	\includegraphics[width=0.49\textwidth]{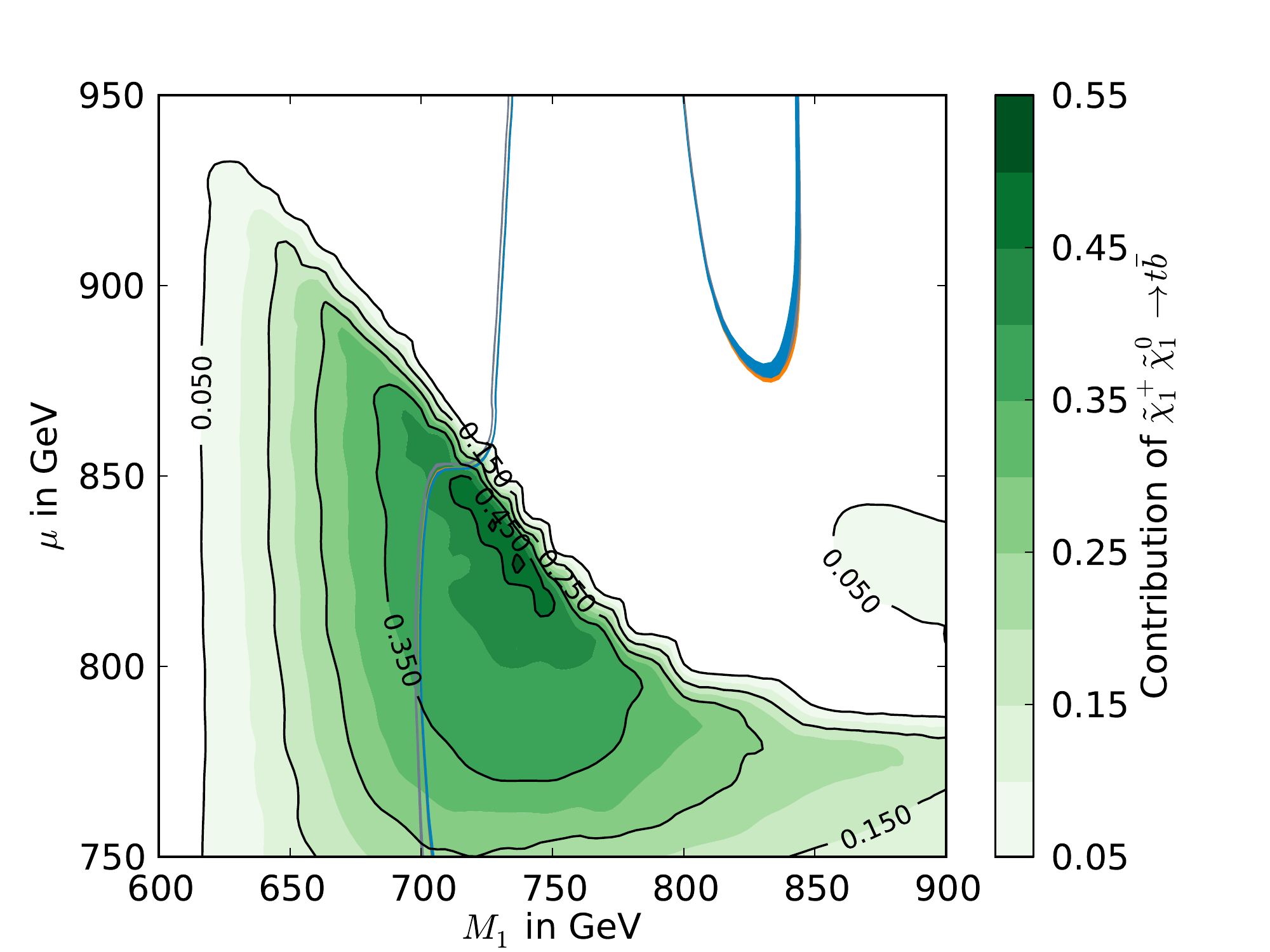}
	\includegraphics[width=0.49\textwidth]{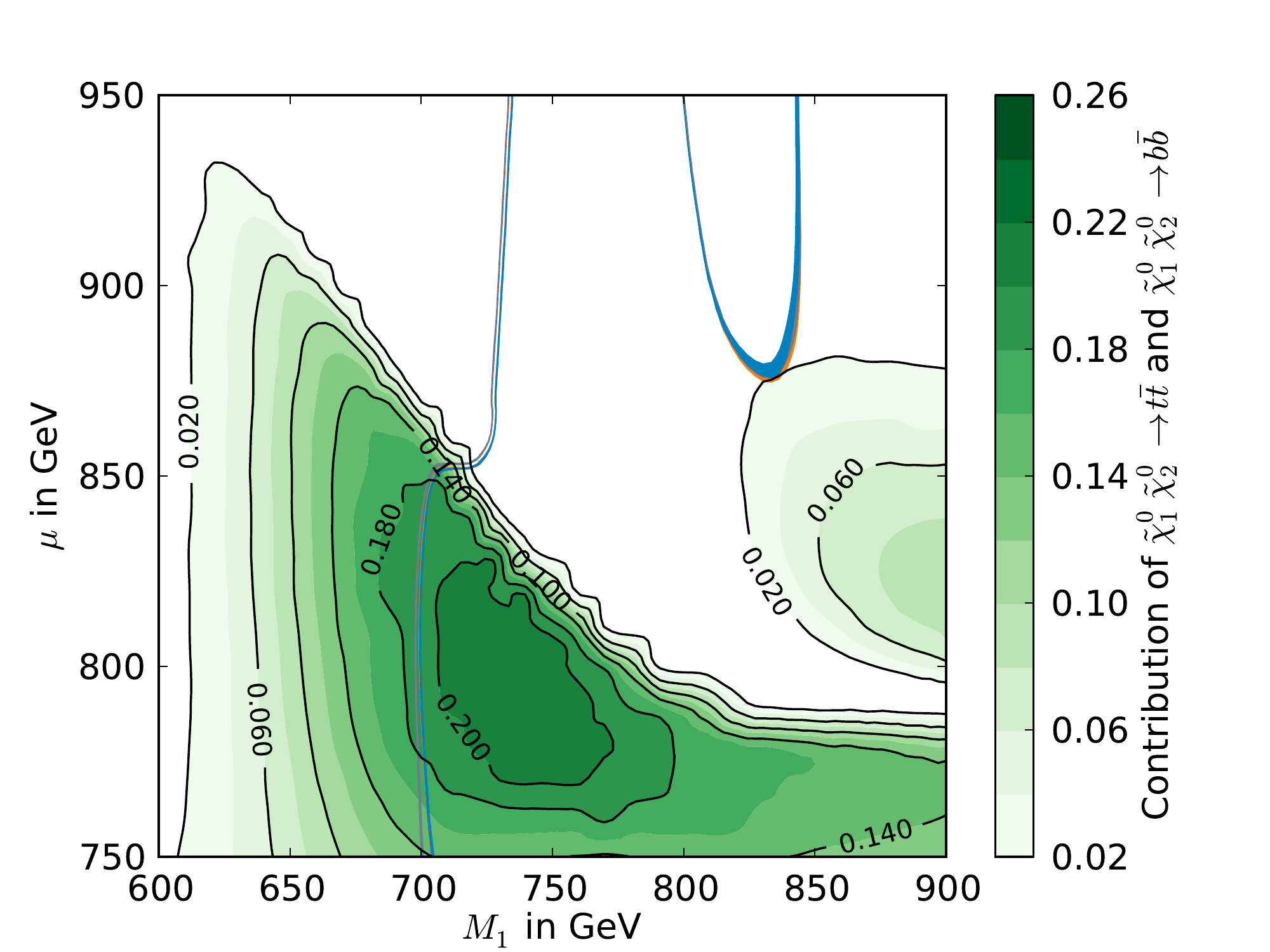}
	\includegraphics[width=0.49\textwidth]{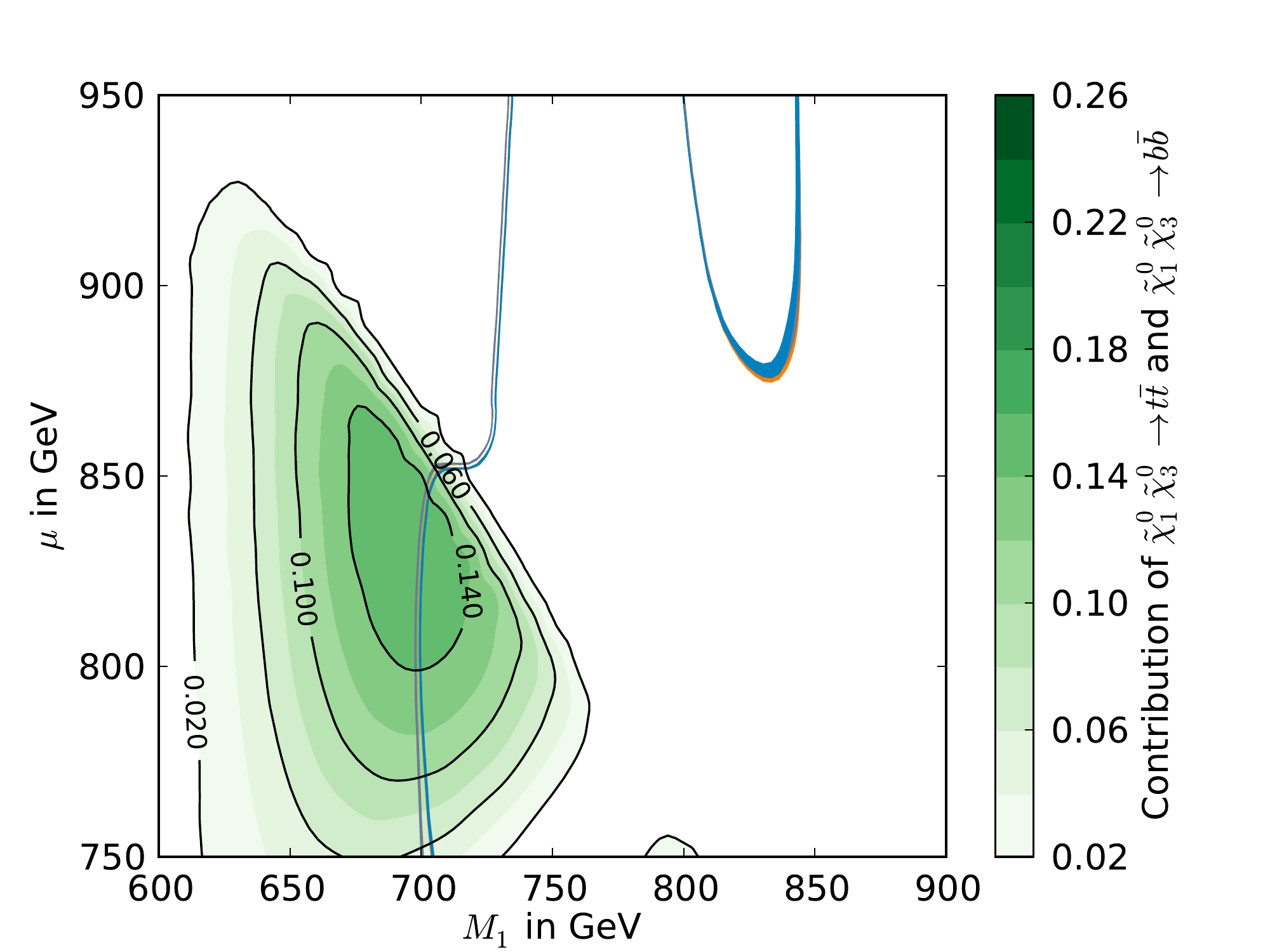}
	\caption{Relative importance of the processes that contribute to the neutralino relic density in the $M_1$--$\mu$ plane surrounding scenario II. The three colored lines represent the part of the parameter space which leads to a neutralino relic density compatible with the Planck limits given in Eq.\ (\ref{Planck}) using the standard \MO\ calculation (orange), our tree-level calculation (grey), and our full one-loop calculation (blue).}
	\label{fig:GC1Plots}
\end{figure*}

\begin{figure*}
	\includegraphics[width=0.49\textwidth]{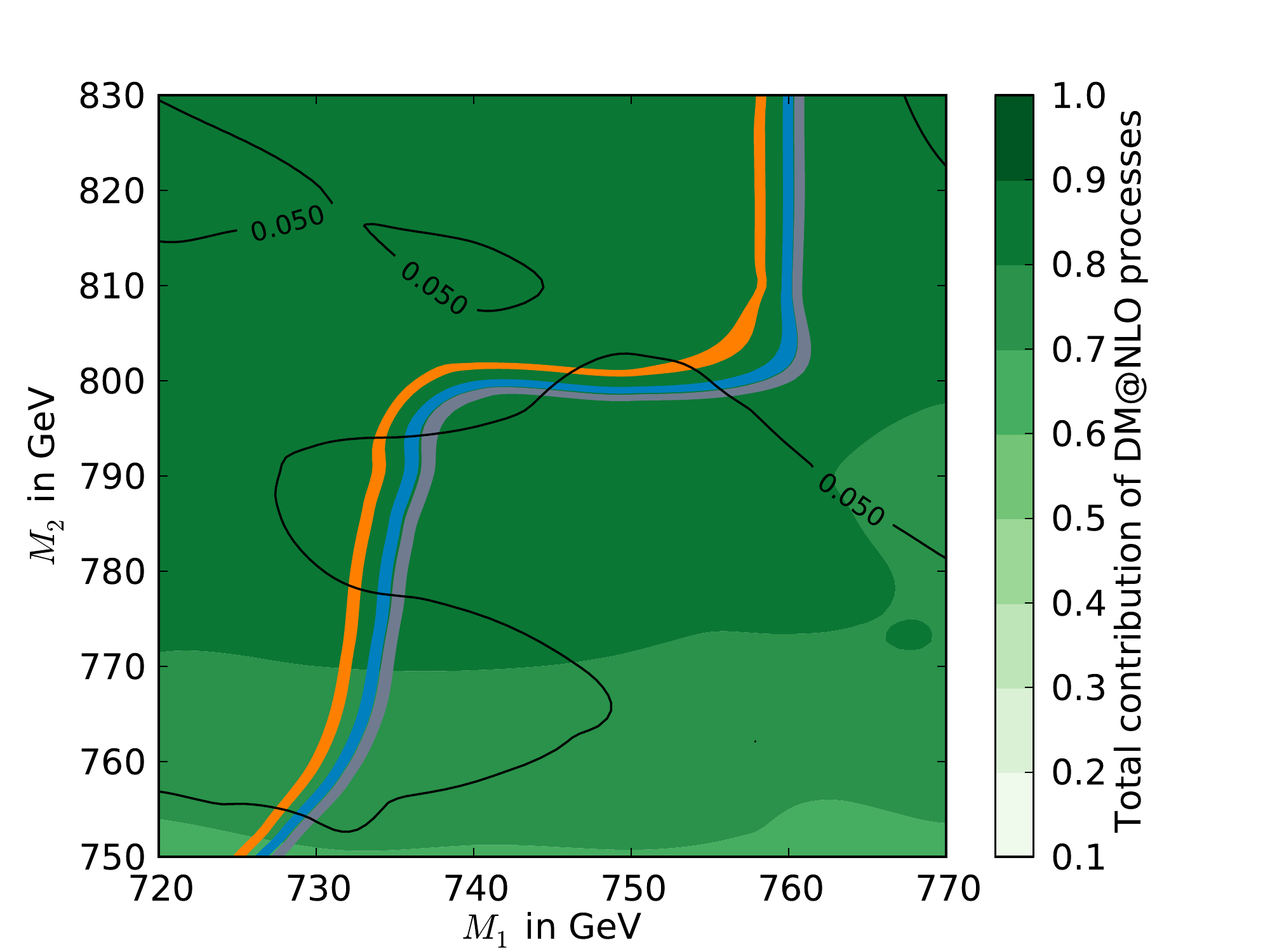}
	\includegraphics[width=0.49\textwidth]{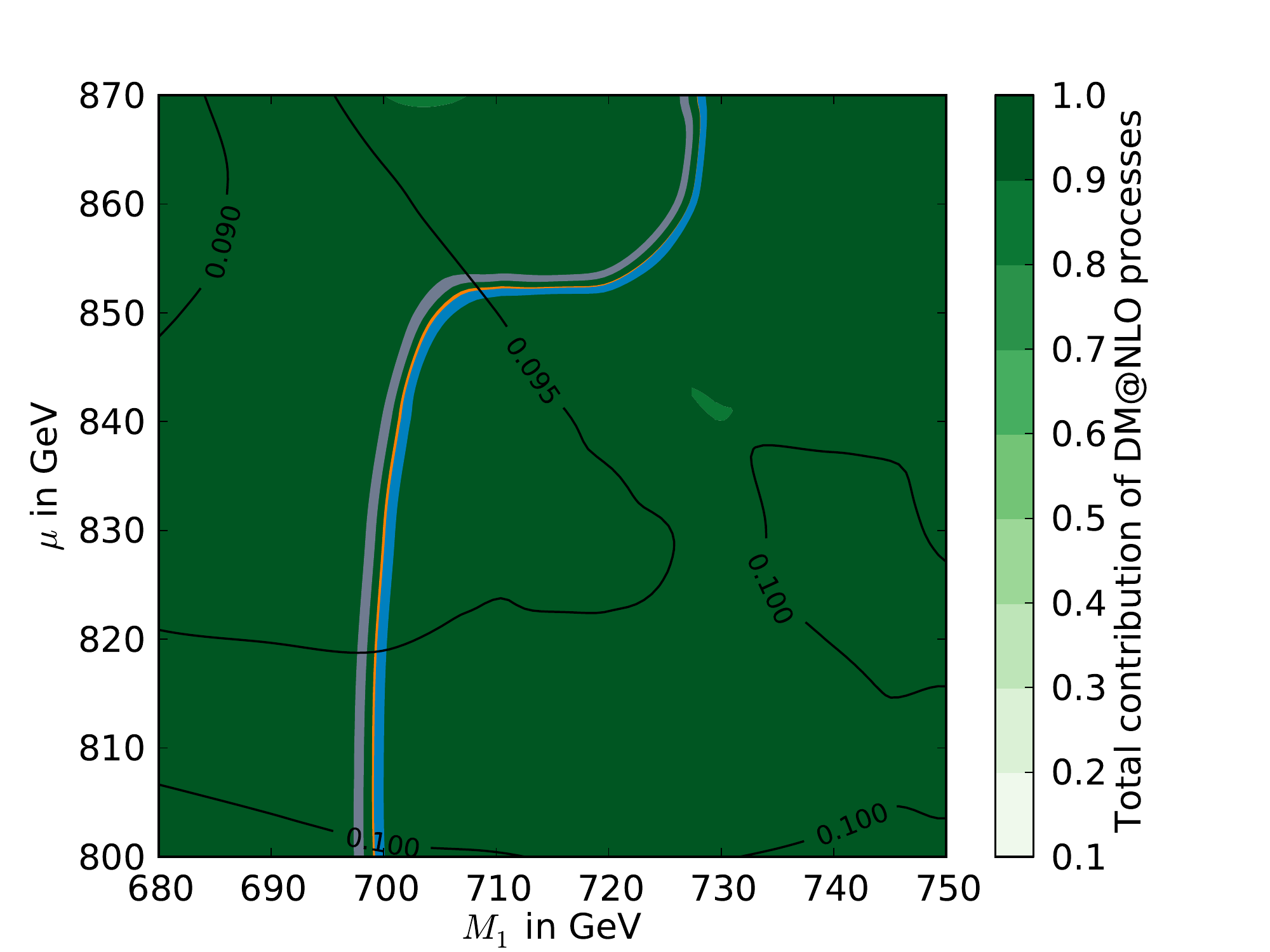}
	\caption{Neutralino relic density in the $M_1$--$M_2$-plane surrounding scenario I (left) and the $M_1$-$\mu$-plane surrounding scenario II (right). The three colored lines represent the part of the parameter space which leads to a neutralino relic density compatible with the Planck limits given in Eq.\ (\ref{Planck}). For the orange line we used the standard \MO\ routine, the grey one corresponds to our tree level calculation, and the blue one represents our full one-loop calculation. The black contour lines denote the relative shift between the tree level and one-loop relic density, i.e.\ $\big|1-\Omega_\chi^{\mathrm{NLO}}/\Omega_\chi^{\mathrm{tree}}\big|$.}
	\label{fig:RelicPlots}
\end{figure*}

The remaining five parts of Fig.\ \ref{fig:G5Plots} illustrate how the total contribution is decomposed into the single channels of our interest. The biggest contribution stems from the annihilation channel $\tilde{\chi}_1^0 \tilde{\chi}_1^0 \rightarrow b\bar{b}$, as can be seen in the upper-right plot. However, the contribution of this channel drops to roughly 20\% in the center-left region, where the total contribution from our processes of interest still accounts for more than 80\%. The main contributions in this part of the parameter space are depicted in the two central plots, namely the co-annihilation channels $\tilde{\chi}_1^0\tilde{\chi}_2^0\rightarrow b\bar{b}$ and $\tilde{\chi}_1^+\tilde{\chi}_1^0\rightarrow t\bar{b}$. These channels contribute up to 24\% and 45\% of the annihilation cross section, respectively, in the cosmologically preferred region. The shape of the two contributions is very similar, which is due to the fact that the particles $\tilde{\chi}_2^0$ and $\tilde{\chi}_1^+$ are almost degenerate in mass (see Tab. \ref{ScenarioProps}). Therefore co-annihilation of $\tilde{\chi}_1^0$ with these two particles becomes important in the same region of the parameter space.

The rather exotic contributions of the channels $\tilde{\chi}_1^+ \tilde{\chi}_2^0 \rightarrow c\bar{s}$ and $\tilde{\chi}_1^+ \tilde{\chi}_2^0 \rightarrow u\bar{d}$ are summed and shown in the lower-left plot. The individual contributions are not shown as they are basically identical. We see that the sum of these channels with light quarks in the final state makes up for 16--18\% in the cosmologically preferred region. The analogous graph for the annihilation channel $\tilde{\chi}_1^+\tilde{\chi}_1^-\rightarrow b\bar{b}$ is shown in the lower-right part of Fig.\ \ref{fig:G5Plots}. This channel constitutes more than 35\% outside and roughly 15\% inside the cosmologically preferred region. Not shown are the contributions from the less important channels $\tilde{\chi}_1^+\tilde{\chi}_1^0\rightarrow c\bar{s}$, $\tilde{\chi}_1^+\tilde{\chi}_1^0\rightarrow u\bar{d}$, $\tilde{\chi}_1^0\tilde{\chi}_1^0\rightarrow t\bar{t}$ ($\sim$ 10\% each), and $\tilde{\chi}_2^0\tilde{\chi}_2^0\rightarrow b\bar{b}$ ($\sim$ 5\%). We emphasize that we correct between two and twelve gaugino annihilation and co-annihilation channels in parallel in the individual scenarios of the $M_1$--$M_2$ plane of Fig.\ \ref{fig:G5Plots} by using the aforementioned filtering mechanism.

Fig.\ \ref{fig:GC1Plots} shows the corresponding graphs for scenario II, where we have performed a scan in the parameters $M_1$ and $\mu$. Here, the total contribution of the gaugino annihilation and co-annihilation processes reaches more than 90\% in a large part of the parameter plane and drops below 10\% only in the upper-right corner, where the $\tilde{t}_1$ becomes degenerate in mass with the lightest neutralino, so that co-annihilation processes between these particles \cite{DMNLO_Stop1, DMNLO_Stop2} and also annihilation processes amongst the stops themselves become dominant.

The largest contributions stem from the channels $\tilde{\chi}_1^0\tilde{\chi}_1^0\rightarrow t\bar{t}$ ($\sim$ 60\%) and $\tilde{\chi}_1^0\tilde{\chi}_1^0\rightarrow b\bar{b}$ ($\sim$ 35\%). Similarly to scenario I, both of these channels are of minor importance in the center-left region, where their contributions drop below 20\% and 10\%, respectively, although the total contribution still accounts for more than 90\%. The most relevant channels in this region are $\tilde{\chi}_+^0\tilde{\chi}_1^0\rightarrow t\bar{b}$ ($\sim$ 45\%), $\tilde{\chi}_1^0\tilde{\chi}_2^0\rightarrow t\bar{t}$ ($\sim$ 13\%), $\tilde{\chi}_1^0\tilde{\chi}_2^0\rightarrow b\bar{b}$ ($\sim$ 7\%), $\tilde{\chi}_1^0\tilde{\chi}_3^0\rightarrow t\bar{t}$ ($\sim$ 9\%) and $\tilde{\chi}_1^0\tilde{\chi}_3^0\rightarrow b\bar{b}$ ($\sim$ 5\%). In contrast to scenario I, $t\bar{t}$ final states are dominant in scenario II. In addition we have contributions from co-annihilation with the third neutralino $\tilde{\chi}_3^0$.

These observations are quite remarkable. Just by investigating the surrounding $M_1$--$M_2$-plane of scenario I, we find that basically all different types of gaugino annihilations and co-annihilations included in Eqs.\ (\ref{NeuNeuAnni}) -- (\ref{ChaChaAnni}) occur in parallel and contribute in a non-negligible way. Even the contributions of light quark final states are sizeable. For scenario II, the surrounding $M_1$--$\mu$ plane also shows a variety of relevant gaugino annihilation and co-annihilation channels. This can be seen as the motivation for our present work. For a precise determination of the neutralino relic density including NLO corrections in $\mathcal{O}(\alpha_s)$ over a broad range of the pMSSM parameter space, it is not sufficient to focus on the annihilation of the LSP into third generation quarks, as it has been previously done. 

In the left-hand plot of Fig.\ \ref{fig:RelicPlots} we show a more detailed view of Fig.\ \ref{fig:G5Plots}. One can see that the three lines, indicating the predictions for the neutralino relic density based on the three calculations, totally separate. This matches our previous observations concerning the upper plots of Fig.\ \ref{fig:CrossSectionPlots}, where the cross sections of the channels $\tilde{\chi}^0_1\tilde{\chi}^+_1\rightarrow t \bar{b}$ and $\tilde{\chi}^0_1\tilde{\chi}^0_2\rightarrow b \bar{b}$ are depicted. As these channels dominate the parameter space around scenario I shown in Fig.\ \ref{fig:RelicPlots} (see also Fig. \ref{fig:G5Plots}), the deviations between the three cross sections propagate through the Boltzmann equation to sizeable deviations in the relic density. In total, our one-loop calculations shift the tree-level relic density by roughly 5\%, whereas the relic density obtained with \MO\ differs by 14\% from our tree-level result.

The right-hand side of Fig.\ \ref{fig:RelicPlots} shows a detail view of Fig.\ \ref{fig:GC1Plots}. Here, the orange and blue lines overlap. This agrees with our results from Sec. \ref{CrossSectionTeil}, where we found that the one-loop corrections of the dominant channels $\tilde{\chi}^0_1\tilde{\chi}^+_1\rightarrow t \bar{b}$ and $\tilde{\chi}^0_1\tilde{\chi}^0_2\rightarrow t \bar{t}$ are well approximated by the effective tree-level calculation from \MO\ (see middle part of Fig. \ref{fig:CrossSectionPlots}). Consequently, the relic density determined with \MO\ agrees with our full one-loop calculation. Nevertheless, we stress that the size of the corrections is larger than previously. The one-loop corrections shift the relic density by roughly 9--10\%.

Let us close the discussion with a comment concerning final states with first and second generation quarks. A scan of the parameter space defined at the beginning of Sec.\ \ref{Pheno} has shown that phenomenologically viable scenarios turn out to be similar to our reference point III, featuring rather heavy squarks of the first and second generation, while those of the third generation are lighter in order to meet the requirement of the 125 GeV Higgs boson. Consequently the NLO corrections to the annihilation cross section into final states with light quarks are relatively small, as can be seen in Fig.\ \ref{fig:CrossSectionPlots} for our scenario III. Thus, this correction to the neutralino relic density will be even less important, since these final states typically account for less than 50\% of the total cross section, the rest being, e.g., annihilation into third-generation quarks. We therefore do not show extensive studies of the relic density for our scenario III. Let us note, however, that in a more general SUSY framework, the situation can be different and the corrections to the (co-)annihilation into light quarks can become numerically relevant. Our numerical code includes these corrections in the most general form, suitable for any MSSM setup.

%% file: conclusion.tex
\section{Conclusion}
\label{Conclusion}

The relic density of a dark matter candidate is determined by its annihilation cross section into Standard Model particles. If there exist further nearly mass-degenerate particles, coannihilation processes become relevant. In this article, we have extended our previous work on neutralino dark matter \cite{DMNLO_AFunnel, DMNLO_mSUGRA, DMNLO_NUHM, DMNLO_Stop1} by studying higher-order corrections in $\alpha_s$ to gaugino annihilation and coannihilation into light and heavy quarks within the eleven-parameter pMSSM for neutralino dark matter. We have calculated all $\mathcal{O}(\alpha_s)$ corrections to the given processes including one-loop $2\rightarrow2$ and tree-level $2\rightarrow3$ amplitudes, leading to a final result that is soft and collinear safe. The numerical integration of the $2\rightarrow3$ processes was rendered finite by making use of the dipole subtraction method for both massive and massless final state quarks. We further improved on the implementation of the bottom quark mass in the $\overline{\rm DR}$-scheme and in addition included the leading NNLO contributions to the bottom quark Yukawa coupling.

We have demonstrated the relevance of these higher-order corrections by investigating three examplary scenarios, which feature a large variety of gaugino annihilation and coannihilation channels occuring in parallel. All three scenarios respect current experimental bounds on the lightest Higgs boson mass and the branching ratio of $b\rightarrow s\gamma$. By including our higher-order corrections, the resulting neutralino relic density was shifted by up to 10\% in comparison to the tree level. This shift is in particular larger than the experimental uncertainty by Planck. Therefore the presented corrections must be taken into account for predicting the neutralino relic density precisely or when extracting SUSY parameters from cosmological measurements.

Further steps in these directions include the completion of all stop coannihilation processes with a gluon final state \cite{DMNLO_Stop2}. We have already encountered the relevance of this processe in the pMSSM scans performed for this work.

%% file: appendix.tex
\section{Dipole formulas}
\label{Dipole:appendix}

Following Ref.\ \cite{Dipole:Dipole0}, we present here further important formulas and definitions. The definitions of the quantities used in Eqs.\ (\ref{Dipole3}) and (\ref{Dipole6}) are:
\begin{equation}
 	\lambda(x,y,z) ~=~ x^2+y^2+z^2-2xy-2xz-2yz ,
\end{equation}
\begin{equation}
 	\tilde{z}_j ~=~ \frac{p_jp_k}{p_ip_k+p_jp_k},
\end{equation}
\begin{equation}
 	y_{ij,k} ~=~ \frac{p_jp_i}{p_ip_k+p_jp_k+p_ip_j},
\end{equation}
\begin{equation}
	v_{ij,k} ~=~ \frac{\sqrt{[2\mu_k^2+(1-\mu_i^2-\mu_j^2-\mu_k^2)(1-y_{ij,k})]^2-4\mu_k^2}}{(1-\mu_i^2-\mu_j^2-\mu_k^2)(1-y_{ij,k})},
\end{equation}
\begin{equation}
 	\mu_i ~=~ \frac{m_i}{\sqrt{s}} ,
\end{equation}
\begin{equation}
 	\mu_{ij}^2 ~=~ \frac{m_i^2+m_j^2}{(p_i+p_j)^2-m_{ij}^2},
\end{equation}
\begin{equation}
 	\tilde{v}_{ij,k} ~=~ \frac{\lambda^{1/2} \big(1,\mu_{ij}^2,\mu_k^2 \big)}{1-\mu_{ij}^2-\mu_k^2}.
\end{equation}

The definitions used in Eq.\ (\ref{Dipole9}) are
\begin{eqnarray}
 	\text{K}_q &~=~& \biggr[ \frac{7}{2}-\frac{\pi^2}{6} \biggr] C_F , \\
 	\gamma_q   & = & \frac{3}{2}C_F,
\end{eqnarray}
as well as 
\begin{equation}
 	\Gamma_q(\mu,m_q;\epsilon)~=~ C_F \left[ \frac{1}{\epsilon} + \frac{1}{2} \ln\frac{m_q^2}{\mu^2}-2 \right]
\end{equation}
for massive quarks and 
\begin{equation}
 	\Gamma_q(\epsilon) ~=~ \frac{1}{\epsilon}\gamma_q
\end{equation}
for massless quarks.

Finally, the singular part of Eq.\ (\ref{Dipole10}) for all possible combinations of emitter-spectator masses are
\begin{widetext}
\begin{eqnarray}
	\mathcal{V}^{(\text{S})}_j(s_{jk},m_j>0,m_k>0;\epsilon) &=& 
		\frac{1}{v_{jk}} \Biggr[ \frac{1}{\epsilon} \log\!\big(\rho\big) -\frac{1}{4} \log^2\!\big(\rho_j^2\big) - 
		\frac{1}{4} \log^2\!\big( \rho_k^2 \big) - \frac{\pi^2}{6}\Biggr] + 
		\frac{1}{v_{jk}} \log\!\big(\rho\big) \log\!\left(\frac{Q_{jk}^2}{s_{jk}}\right), ~~~ \\
 	\mathcal{V}^{(\text{S})}_j(s_{jk},m_j>0,0;\epsilon) &=& 
		\frac{1}{2\epsilon^2}+\frac{1}{2\epsilon} \log\!\left(\frac{m_j^2}{s_{jk}} \right) -
		\frac{1}{4}\log^2\!\left(\frac{m^2_j}{s_{jk}}\right) - \frac{\pi^2}{12} -
		\frac{1}{2}\log\!\left(\frac{m_j^2}{s_{jk}}\right) \log\!\left(\frac{s_{jk}}{Q_{jk}^2}\right) \nonumber \\
	& & -\frac{1}{2} \log\!\left(\frac{m_j^2}{Q^2_{jk}}\right) \log\!\left(\frac{s_{jk}}{Q^2_{jk}}\right), \\
 	\mathcal{V}^{(\text{S})}_j(s_{jk},0,0;\epsilon) &=& \frac{1}{\epsilon^2} ,
\end{eqnarray}
and the non-singular part takes the form
\begin{eqnarray}
	\nonumber \mathcal{V}^{(\text{NS})}_j(s_{jk},m_j\!>\!0,m_k\!>\!0) & ~=~ & 
		\frac{\gamma_q}{\bold{T}_q^2} \log\!\, \biggr(\frac{s_{jk}}{Q_{jk}^2}\biggr) 
		- \frac{m_k}{Q_{jk}-m_k} + \frac{2 m_k(2m_k-Q_{jk})}{s_{jk}} + \frac{\pi^2}{2} \nonumber \\
	& & + \frac{1}{v_{jk}}\left[ \log\!\big(\rho^2\big) \log\! \big(1+\rho^2\big) + 2\text{Li}_2\big(\rho^2\big) - 
		\text{Li}_2 \big(1-\rho_j^2\big) - \text{Li}_2 \big(1-\rho_k^2\big) -\frac{\pi^2}{6} \right] \nonumber \\ 
	& &	+ \log\!\left( \frac{Q_{jk}-m_k}{Q_{jk}} \right) - 
		2 \log\!\left( \frac{(Q_{jk}-m_k)^2-m_j^2}{Q_{jk}^2} \right) - 
		\frac{2m_j^2}{s_{jk}} \log\!\left( \frac{m_j}{Q_{jk}-m_k} \right), ~~~~  
\eea
\bea
	\mathcal{V}^{(\text{NS})}_j(s_{jk},m_j>0,0) &=&
		\frac{\gamma_q}{\bold{T}_q^2} \log\!\left(\frac{s_{jk}}{Q_{jk}^2}\right) + 
		\frac{\pi^2}{6}-\text{Li}_2\left(\frac{s_{jk}}{Q_{jk}^2}\right) -
	 	2\,\log\! \left(\frac{s_{jk}}{Q^2_{jk}}\right) - \frac{m^2_j}{s_{jk}}\log\! \left(\frac{m_j^2}{Q^2_{jk}}\right),\\
	\mathcal{V}^{(\text{NS})}_j(s_{jk},0,m_k>0) &=&
		\frac{\gamma_q}{\bold{T}_q^2} \left[\log\!\left(\frac{s_{jk}}{Q_{jk}^2}\right) - 
		2\,\log\!\left(\frac{Q_{jk}-m_k}{Q_{jk}}\right)- 
		\frac{2m_k}{Q_{jk}+m_k}\right]+\frac{\pi^2}{6} - 
		\text{Li}_2\left(\frac{s_{jk}}{Q_{jk}^2}\right),\\
	\mathcal{V}^{(\text{NS})}_j(s_{jk},0,0) &=& 0.
\end{eqnarray}
\end{widetext}
Here we have used
\begin{equation}
 \rho_n(\omega_j,\omega_k) ~=~
 	\sqrt{\frac{1-v_{j,k}+2\omega_n/(1-\omega_j-\omega_k)}{1+v_{j,k}+2\omega_n/(1-\omega_j-\omega_k)}}
\end{equation}
for $n=j,k$ and
\begin{equation}
 \rho ~=~ \sqrt{\frac{1-v_{j,k}}{1+v_{j,k}}}
\end{equation}
with
\beq
	\omega_n ~=~ \frac{m_n^2}{Q_{jk}^2},
	\eeq
	\beq
 	Q_{jk}  = \sqrt{Q_{jk}^2}=\sqrt{s_{jk}+m_j^2+m_k^2},
 	\eeq
 	\beq
	v_{j,k}  = \frac{\sqrt{\lambda((p_j+p_k)^2,p^2_j,p^2_k)}}{(p_j+p_k)^2-p_j^2-p_k^2}.
	\eeq

%% file: bib.tex
\bibliographystyle{apsrev}